\documentclass[a4paper,11pt]{article}
\usepackage{amsmath,amssymb,mathrsfs,cmbright}
\usepackage[top=3cm,bottom=3cm,left=3cm,right=3cm]{geometry}
\usepackage{color}
\usepackage[T1]{fontenc}
\usepackage{hyperref}
\hypersetup{colorlinks=true}
\usepackage{cite}
\date{}
\begin{document}
\author{Burak Tevfik Kaynak${^\dag}$ and O. Teoman Turgut$^\ddag$ \\ Department of Physics, Bo\u{g}azi\c{c}i University \\ 34342 Bebek, Istanbul, Turkey \\ $^\dag$burak.kaynak@boun.edu.tr, $^\ddag$turgutte@boun.edu.tr}
\title{\bf Compact submanifolds supporting singular interactions}
\maketitle
\begin{abstract}
A quantum particle moving under the influence of singular interactions on embedded surfaces furnish an interesting example from the spectral  point of view. In  these problems, the possible occurrence  of a bound-state is perhaps the most important aspect. Such systems can be introduced as quadratic forms and generically they do not require renormalization. Yet an alternative path through the resolvent is also beneficial to study various properties. In the present work, we address these issues for compact surfaces embedded in a class of ambient manifolds. We discover that there is an exact bound state solution written in terms of the heat kernel of the ambient manifold for a range of coupling strengths. Moreover, we develop techniques to estimate bounds on the ground state energy when several surfaces,  each  of which admits a bound state solution, coexist.     
\end{abstract}
\section{Introduction}\label{s1}
In the present work, we will study singular interactions supported on embedded surfaces in some three dimensional ambient manifolds, whose features would be specified. These problems can be thought of as delta function potentials concentrated on some surfaces, therefore they should be carefully defined. Since we will consider codimension one case, one does not expect a singularity, which would require renormalization. The situation is indeed different for point interactions on two or three dimensional manifolds in this respect. Such class of problems forms a natural laboratory to understand some aspects of nonperturbative renormalization. These problems are thoroughly investigated in Refs.~\cite{al1,al2}, and the references therein.

With the advances in mesoscopic systems, the possibility of trapping particles in nontrivial geometries appear. An idealization of these systems would be to model them as some singular potentials concentrated on submanifolds with different codimensions. In the pioneering works~\cite{ex0,ex1,ex2,ex3,ex4,ex5}, the authors define such Hamiltonians as quadratic forms, and investigate several interesting aspects of the resulting spectrum. They show that under certain conditions curvature induces a bound-state in the spectrum. The authors in Ref.~\cite{ex3} study the infinite coupling limit of such trapping potentials, which could be considered as a quantum constraint system. In Ref.~\cite{db}, the bound-state problem of the Schr\"{o}dinger equation for a radially symmetric attractive potential with any finite number of Dirac delta functions is studied. The authors in Ref.~\cite{lin} generalize the result of the work~\cite{ex1} to hypersurfaces in $\mathbb{R}^{n+1}$ under various geometric conditions.   

In Refs.~\cite{bt1,bt2}, we studied the singular interactions supported on embedded curves in Riemannian manifolds through a resolvent formula and heat kernel techniques. Those problems require renormalization, therefore the resolvent approach becomes the most efficient and natural one. In this paper, we will employ the resolvent approach as well as the variational methods, thinking of the Hamiltonian as a quadratic form, to understand the bound-state spectrum in the case of compact submanifolds embedded in Riemannian manifolds with a few mild conditions imposed. Due to the geometric nature of this problem, the heat kernel becomes an essential part of our discussion, and allows us to propose an explicit solution of the bound-state wave function. Using intrinsic geometry of the submanifold as well as geometric aspects of the ambient manifold, we found limits of the strength of the interaction, affirming the existence of bound-states. Although this problem does not require a renormalization, contrary to the aforementioned codimension two cases, this scheme still provides easier paths to study certain features of the model. Moreover, if one would like to analyze mixed systems such as surfaces coexisted with points or curves, then this approach is the most advantageous one to define and study the system.          

The plan of the paper is as follows: in Section~\ref{s2}, we will start constructing the model as a limit of regular potentials through a resolvent formula. For a single submanifold we will define the model as a quadratic form, and then propose an exact bound-state solution by the variational method whereby we obtain the principal operator for a collection of submanifolds provided each one of them admits a bound-state on its own. In Section~\ref{s3}, our vigorous pursuit is to determine the range of the interaction strength so as to affirm occurrence of a bound-state solution as formulated in Section~\ref{s2}. In Section~\ref{s4}, the fact that the theory is finite will be shown by demonstrating the finiteness of the principal matrix. In Section~\ref{intermezzo}, we will generalize the variational approach to a collection of submanifolds. In Section~\ref{s6}, we will demonstrate that the ground state energy is bounded from below after holding a short discussion about the uniqueness of the ground state. Afterwards, possible issues which are worthy of remark will be addressed in light of simple, albeit important, examples. In Section~\ref{s7}, some comments on more general cases will be made.
\section{Formulation of the model by a bound-state Hamiltonian}\label{s2}
In this section, we will study the construction of the model, which  describes a non-relativistic particle interacting with a singular potential concentrated on distinct surfaces while moving on a curved background. Our main interest is the bound-state structure of this model problem. However, we would like to refer briefly to some notions related to geometric  settings~\cite{wil,klin} on which the model is built before going into details.
 
Let $(\mathcal{M},\tilde{g})$ be a 3-dimensional Riemannian manifold, admitting a lower bound on its Ricci curvature, and $(\Sigma,g)$ be a compact orientable 2-dimensional Riemannian manifold, whose sectional curvature is subjected to have both upper and lower bounds. Let $\iota$ be an inclusion map $\iota \,:\,\Sigma \hookrightarrow \mathcal{M}$, assumed to be an isometric embedding, then $\Sigma$ is said to be an isometrically embedded submanifold of the ambient manifold $\mathcal{M}$. Due to this embedding, the Riemannian metric $\tilde{g}$ on $\mathcal{M}$ induces the Riemannian metric $g=\iota^*\tilde g $ on $\Sigma$. Henceforth, whenever we refer to the points on the ambient manifold $\mathcal{M}$ we use the symbol $\tilde x$, and similarly for tensors on the ambient manifold,  and for objects on the submanifold we will not have a tilde. That the inclusion map is an embedding excludes the possibility of self-intersections of the submanifolds. Moreover, we restrict ourselves with the non-intersecting submanifolds whenever a collection of submanifolds is under consideration.

In this geometric framework, we will develop the model as a generalized Schr\"{o}dinger operator with a singular interaction, whose support is an isometrically embedded 2-dimensional compact submanifold $\Sigma$ of a 3-dimensional ambient manifold $\mathcal{M}$. The generalized Schr\"{o}dinger equation for the model with a single submanifold is given by
\begin{align}\label{sch}
-\frac{\hbar^2}{2 m} \nabla_{\tilde{g}}^2 \psi(\tilde x) - \frac{\lambda}{V(\Sigma)} \int_\Sigma d_g \mu (x') \, \delta_{\tilde{g}}(\tilde x,x') \int_\Sigma d_g \mu (x'') \, \psi(x'') &= E \psi(\tilde x) \,,
\end{align}
where $\nabla^2_{\tilde{g}}$ is the generalized Laplacian, $V(\Sigma)$ is the volume(area) of the submanifold, $ d_g \mu (x')$ is the measure on the submanifold at the point $x' \in \Sigma$, which comes from the pull-back of the Riemannian volume element on $\mathcal{M}$  through the embedding.  $\delta_{\tilde{g}}(\tilde x,\tilde x')$ is the generalized Dirac-delta function and $\lambda$ is the coupling constant, describing the interaction strength. Henceforward, we will omit the metric from the measure. 
       
Following Refs.~\cite{bt1,bt2}, we will rewrite the interaction as a projection operator for the collection of submanifolds, each is labeled by $i$,  as the distinct submanifolds are non-intersecting. For this purpose, the following family of functions supported on submanifolds, each thereof is also labeled by $i$,  are introduced,
\begin{align}
\Gamma_i^\epsilon(\tilde x) &= \int_\Sigma d \mu(x') \, K_{\epsilon/2} (\tilde x,x') \,,
\end{align}   
$K_{\epsilon/2} (\tilde x,\tilde x')$ being the heat kernel. It is the positive fundamental solution of the heat equation for the Laplace-Beltrami operator, i.e.,
\begin{align}
- \frac{\hbar^2}{2 m} \nabla_{\tilde g}^2 K_t(\tilde x,\tilde x') &= - \hbar \frac{\partial}{\partial t} K_t(\tilde x,\tilde x') \,.
\end{align}
Furthermore, the heat kernel enjoys the semi-group property, which is
\begin{align}
\int_\mathcal{M} d\tilde \mu(\tilde x'') K_t(\tilde x,\tilde x'') K_{t'}(\tilde x'',\tilde x') &= K_{t+t'}(\tilde x,\tilde x') \,,
\end{align}
where $ d \tilde \mu (\tilde x)$ is the measure on the ambient manifold at the point $\tilde x \in \mathcal{M}$. This property allows us to calculate the  inner product of these functions:
\begin{align}
\langle \Gamma_i^\epsilon \vert \Gamma_i^\epsilon \rangle &= \int_{\Sigma_i \times \Sigma_i} d \mu(x) d \mu(x') \, K_\epsilon (x,x') \,,
\end{align} 
which is finite. $\Gamma_i^\epsilon$  becomes a Dirac-Delta function supported on $i$-th submanifold after  taking the limit $\epsilon \rightarrow 0^+$. In terms of the free Hamiltonian and the projection operators defined above, the Schr\"{o}dinger operator with singular interactions supported on a collection of submanifolds,  takes the following form,
\begin{align}\label{sc}
(H_0 - E) \vert \psi \rangle &= \sum_i \frac{\lambda_i}{V(\Sigma_i)} \vert \Gamma_i^\epsilon \rangle \langle \Gamma_i^\epsilon \vert \psi \rangle \,.
\end{align}
Since our interest is mainly on the bound-state structure of the model, it is more natural to  obtain the resolvent of the interacting Hamiltonian instead of using the formal delta functions directly. In order to determine the full resolvent, we need to solve $\langle \tilde{\Gamma}_i^\epsilon \vert \psi \rangle$ in the following expression, in which $\sqrt{\lambda_i/V(\Sigma_i)}$ is absorbed in the definition of $\Gamma_i^\epsilon$,
\begin{align}\label{psi}
\vert \psi \rangle &= \sum_i (H_0 - E)^{-1} \vert \tilde{\Gamma}_i^\epsilon \rangle \langle \tilde{\Gamma}_i^\epsilon \vert \psi \rangle + (H_0 - E)^{-1} \vert \varphi \rangle \,. 
\end{align}
We can easily show that $\langle \tilde{\Gamma}_i^\epsilon \vert \psi \rangle$ is given by the following matrix equation,
\begin{align}\label{gpsi}
\langle \tilde{\Gamma}_i^\epsilon \vert \psi \rangle &= \sum_j \left[ \frac{1}{1-\langle \tilde{\Gamma}^\epsilon \vert (H_0-E)^{-1} \vert \tilde{\Gamma}^\epsilon \rangle} \right]_{ij} \langle \tilde{\Gamma}_j^\epsilon \vert (H_0-E)^{-1} \vert \varphi \rangle \,.
\end{align} 
Plugging Eq.~(\ref{gpsi}) into Eq.~(\ref{psi}), and rescaling back by $\sqrt{\lambda_i/V(\Sigma_i)}$ thereafter put the full resolvent into the following form,
\begin{align}\label{fg}
\frac{1}{H-E} &= \frac{1}{H_0-E} + \frac{1}{\sqrt{V(\Sigma_i)V(\Sigma_j)}} \frac{1}{H_0-E} \vert \Gamma_i^\epsilon \rangle \frac{1}{\Phi^\epsilon_{ij}} \langle \Gamma_j^\epsilon \vert \frac{1}{H_0-E} \,, 
\end{align}
where there are summations over repeated indices. The operator $\Phi^\epsilon_{ij}$ refers to the so-called principal matrix (operator),  essentially all  the information of bound-states is kept in this matrix (also called the Krein function in the mathematical literature). A  general approach to such singular problems in this perspective is presented in~\cite{pos}. It is the only matrix that has an inverse in the resolvent formula besides the free Hamiltonian. In the case of compact manifolds, the Laplacian is self-adjoint and has a spectrum which consists of eigenvalues only.  On a complete non-compact manifold, the Laplacian is a positive essentially-self adjoint operator, nevertheless its spectrum may be very complicated. Its essential spectrum (if it exists) is contained in the interval $[0,\infty)$. From a technical point of view the derivation below does not require anything further than this. Nevertheless, if we have gaps in the spectrum, then the search for bound states may become more involved. Since the appearance of an isolated eigenvalue  in this gap as a result of the interaction should also be investigated. Therefore, to avoid complications which may arise due to possible gaps in the essential spectrum, in this work \emph{the noncompact manifolds we consider  will be subjected to the condition given in the recent work by Lu and Zhou\cite{zhou}, i.e. the Ricci tensor is asymptotically nonnegative}. The reader is invited to consult to  this work for a detailed discussion of this condition and its relation to previous investigations.  As a result of this study one knows that for such manifolds, the Laplacian   has essential  spectrum equal to the interval $[0,\infty)$. In the case of compact manifolds this issue naturally arises again since the full spectrum is discrete. In this case as well, to be consistent with the noncompact case we will assume that bound state means an eigenvalue below the bottom of the spectrum. This assumption can further be supported by considering the possibility of taking large volume limits of compact manifolds without turning the bound state energy into a possible resonance. Therefore, in the compact and the specified noncompact cases, the zeros of the principal  matrix  on the negative axis should correspond to the poles of the resolvent, and must determine bound-states. 

It can be shown that the principal matrix is  given by
\begin{align}\label{ph}
\Phi_{ij}^\epsilon(E) &= \left\{ \begin{array}{l} \frac{1}{\lambda_i} - \frac{1}{V(\Sigma_i)} \langle \Gamma_i^\epsilon \vert (H_0 - E)^{-1} \vert \Gamma_i^\epsilon \rangle \,, \\ - \frac{1}{\sqrt{V(\Sigma_i) V(\Sigma_j)}} \langle \Gamma_i^\epsilon \vert (H_0 - E)^{-1} \vert \Gamma_j^\epsilon \rangle \,. \end{array} \right.
\end{align}   
The free resolvent is none other than the integral of the heat kernel over its time parameter,
\begin{align}\label{inv}
\langle \tilde x \vert\frac{1}{H_0-E} \vert \tilde x' \rangle &= \int_0^\infty \frac{d t}{\hbar} \, e^{E t/ \hbar} K_t(\tilde x,\tilde x') \,.
\end{align}
After plugging Eq.~(\ref{inv}) into Eq.~(\ref{ph}) and using the semi-group property of the heat kernel, the diagonal part of the principal matrix is found to be
\begin{align}\label{onp}
\Phi_{ii}^\epsilon(E) &= \frac{1}{\lambda_i}-\frac{1}{V(\Sigma_i)} \iint_{\Sigma_i \times \Sigma_i} d \mu (x)  d \mu (x') \int_0^\infty \frac{d t}{\hbar} \, e^{E (t-\epsilon) / \hbar} K_t(x,x') \,.
\end{align}
Whereas the models studied in Refs.~\cite{bt1,bt2} require renormalization to be well-defined, {\it this problem does not need to be renormalized}, which allows us to safely take the limit $\epsilon \rightarrow 0^+$. However, we will explicitly demonstrate that it is a finite theory on the forthcoming pages. Although the expression in Eq.~(\ref{onp}) is valid for $E<0$, it can be extended to the case where $E$ belongs to the complex plane by a proper analytical continuation, which we will elaborate below. Taking the aforesaid limit and replacing $E$ by $-\nu^2$ for convenience (searching for bound state solutions) allows us to obtain the following form for the principal matrix,
\begin{align}\label{phi0}
\Phi_{ij}(-\nu^2) &= \left\{ \begin{array}{l} \frac{1}{\lambda_i}-\frac{1}{V(\Sigma_i)} \iint_{\Sigma_i \times \Sigma_i} d \mu (x) d \mu (x') \int_0^\infty \frac{d t}{\hbar} \, e^{-\nu^2 t / \hbar} K_t(x,x') \,, \\ -\frac{1}{\sqrt{V(\Sigma_i)V(\Sigma_j)}} \iint_{\Sigma_i \times \Sigma_j} d \mu (x)  d \mu (x') \int_0^\infty \frac{d t}{\hbar} \, e^{-\nu^2 t / \hbar} K_t(x,x') \,. \end{array} \right.
\end{align}

One of the important questions here to ask is that under which conditions one can observe a bound-state or more precisely a discrete spectrum, and the answer to this question for a variety of special geometries are addressed extensively in Refs.\cite{ex1,ex2,ex3,ex4,ex5,lin} by different techniques. In this work, as mentioned in Section~\ref{s1}, we will follow a completely different approach. At first we will search for the ground state wave function for a given bound-state energy by the variational method. Afterwards, we will look for a solution of the variational equation for the ground state energy, and will show that one can \textit{exactly} solve this equation provided the coupling constant is defined in a particular form. This determines the value of the coupling constant for a given bound-state energy, thereby allowing us to construct the Hamiltonian whose spectrum contains a discrete subset. The variational principle assures that the proposed wave function corresponds to the ground state. 

Let us start with the following choice of ansatz for the ground state wave function of a model consisting of a single submanifold for simplicity,
\begin{align}\label{w}
\psi_\alpha(\tilde x) &= \int_\Sigma d \mu(x') \int_0^\infty \frac{d t}{\hbar} \, e^{-\alpha t/\hbar} K_t(\tilde x,x') \,.
\end{align}
Since the normalized wave function is needed for the variational principle, first we should calculate the normalization of this wave function.
\begin{align}
Z(\alpha) &= \int_\mathcal{M} d \tilde \mu (\tilde x) \, \vert \psi_\alpha(\tilde x) \vert^2 \,,
\end{align}                    
$\alpha$ being the variational parameter. Note that for any $\alpha$ this wave function is normalizable since $\Sigma$ is compact. After plugging our ansatz~(\ref{w}) into the expression above, we obtain
\begin{align}
Z(\alpha) &= \int_\mathcal{M} d \tilde \mu(\tilde x'') \iint_{\Sigma \times \Sigma} d \mu(x) d \mu(x') \int_0^\infty \frac{d u}{\hbar} \int_0^\infty \frac{d v}{\hbar} \, e^{- \alpha (u + v)/ \hbar} K_u(x,\tilde x'')  K_v(\tilde x'',x')\,.
\end{align}
The normalization of our trial wave function can be turned into a simple form via  using the semi-group property of the heat kernel and substituting the new time variables $t=u+v,t'=u-v$ in the integrals, and by evaluating the resulting $t'$ integral. Therefore, it reads  
\begin{align}\label{z}
Z(\alpha) &= \iint_{\Sigma \times \Sigma} d \mu(x) d \mu(x') \int_0^\infty \frac{d t}{\hbar} \, \frac{t}{\hbar} e^{- \alpha t / \hbar} K_t(x,x') \,.
\end{align}
The expectation value of the Hamiltonian as a function of the variational parameter is given by
\begin{align}
E(\alpha) &= \frac{1}{Z(\alpha)} \int_\mathcal{M} d \tilde \mu(\tilde x) \, \psi_\alpha(\tilde x) \left[ -\frac{\hbar^2}{2 m} \nabla_{\tilde g}^2 \psi_\alpha (\tilde x) - \frac{\lambda}{V(\Sigma)} \int_\Sigma d \mu (x') \, \delta(\tilde x,x') \int_\Sigma d \mu (x'') \, \psi_\alpha(\tilde x'') \right] \,. 
\end{align}
After plugging our choice of ansatz into the equation above, the Laplacian in the first term in the parenthesis can be replaced by the time derivative since the heat kernel is the positive fundamental solution of the heat equation. A partial integration makes this term free of the derivative operator, as a result the expectation value of the Hamiltonian is obtained as
\begin{align}\label{en}
E(\alpha) &= \frac{1}{Z(\alpha)} \int_\Sigma d \mu(x) \, \psi_\alpha(x) - \alpha - \frac{\lambda}{V(\Sigma)} \frac{1}{Z(\alpha)} \left[ \int_\Sigma d \mu(x) \, \psi_\alpha(x) \right]^2 \,.
\end{align}
We note that the integral of the wave-function over the submanifold is actually finite, as we will see in more detail in Section~\ref{s4}.
Now, we need to find the value of the variational parameter $\alpha$ at which the first partial derivative of $E(\alpha)$ with respect to $\alpha$ is equal to zero, that is,
\begin{align}
\frac{\partial E(\alpha)}{\partial \alpha} \Big \vert_{\alpha = \alpha_*} &= 0 \,.
\end{align}
This gives none other than an extremum of the energy functional above. The first derivative is given by
\begin{align}\label{den}
\frac{\partial E(\alpha)}{\partial \alpha} &= \left[\frac{\partial}{\partial \alpha} \frac{1}{Z(\alpha)} \right] \left( \int_\Sigma d \mu(x) \, \psi_\alpha(x) - \frac{\lambda}{V(\Sigma)} \left[ \int_\Sigma d \mu(x) \, \psi_\alpha(x) \right]^2 \right) \nonumber \\
& \quad -2 + \frac{2 \lambda}{V(\Sigma)} \int_\Sigma d \mu(x) \, \psi_\alpha(x) \,.
\end{align}
It is obvious that the choice of the coupling constant below for a special value of the variational parameter $\alpha=\alpha_*$ not only makes the right-hand side of Eq.~(\ref{den}) equal to zero but also allows us to solve Eqs.~(\ref{sch}) and~(\ref{en}) \textit{exactly},
\begin{align}\label{la}
\frac{1}{\lambda} &= \frac{1}{V(\Sigma)} \int_\Sigma d \mu(x)\psi_{\alpha_*}(x) \,.
\end{align}
If we utilize the ansatz for the ground state wave function, given by Eq.~(\ref{w}), in the choice for the coupling constant above, then we can obtain how the inverse coupling constant should be defined as a double-integral which is taken over the points on the submanifold while these points are connected by the heat kernel of the ambient manifold. In that way, the coupling constant does not only depend on the geometric features of the submanifold, but also depends on the geometric structure of the ambient manifold. Since Eq.~(\ref{la}) is an integration of the wave function over the submanifold, we express this wave function as a convolution of a Dirac-delta function with the original wave function (which is a function of the points in the ambient manifold). Therefore, we have   
\begin{align}
\frac{1}{\lambda} &= \frac{1}{V(\Sigma)} \iint_{\Sigma \times \Sigma} d \mu(x) d \mu(x') \int_{\mathcal{M}} d \mu (\tilde{x}) \delta_{\tilde{g}}(\tilde{x},x) \int_0^\infty \frac{d t}{\hbar} \, e^{- \alpha_* t / \hbar} K_t(\tilde{x},x')\,.
\end{align}
The inverse coupling constant, after integrating over the ambient manifold, reads        
\begin{align}\label{lam}
\frac{1}{\lambda} &= \frac{1}{V(\Sigma)} \iint_{\Sigma \times \Sigma} d \mu(x) d \mu(x') \int_0^\infty \frac{d t}{\hbar} \, e^{- \alpha_* t / \hbar} K_t(x,x') \,.
\end{align}
We will choose $\alpha_*$ as the absolute value of the bound-state energy $E_b=-\nu_*^2$ such that Eq.~(\ref{lam}) is satisfied. On the other hand, it is not sufficient to have the zero of the first derivative of $E(\alpha)$ at $\alpha_*$ so as to affirm that $\alpha_*$ is the minimum. Therefore, we should observe that the second partial derivative of $E(\alpha)$ with respect to $\alpha$ is positive at the value chosen above. It is given by
\begin{align}\label{2nd}
\frac{\partial^2 E(\alpha)}{\partial \alpha^2} \Big \vert_{\alpha=\alpha_*} &= \iint_{\Sigma \times \Sigma} d \mu(x) d \mu(x') \int_0^\infty \frac{d t}{\hbar} \, e^{-\alpha_* t / \hbar} K_t(x,x') \left( \frac{1}{Z(\alpha_*)} \frac{t^2}{\hbar^2} - \frac{2 \lambda}{V(\Sigma)} \frac{t}{\hbar} \right) \,.
\end{align}
To accomplish this task, let us reorganize Eq.~(\ref{2nd}) using Eqs.~(\ref{z}) and~(\ref{lam})  to arrive the following form,
\begin{align}\label{zz}
Z(\alpha_*) \frac{\partial^2 E(\alpha_*)}{\partial \alpha^2} &= \iint_{\Sigma \times \Sigma} d \mu(x) d \mu(x') \int_0^\infty \frac{d t}{\hbar} \, \frac{t^2}{\hbar^2} e^{-\alpha_* t / \hbar} K_t(x,x') \nonumber \\
& \qquad - 2 \left[ \iint_{\Sigma \times \Sigma} d \mu(x) d \mu(x') \int_0^\infty \frac{d t}{\hbar} \, e^{-\alpha_* t / \hbar} K_t(x,x') \right]^{-1} \nonumber \\
& \qquad \times \left[ \iint_{\Sigma \times \Sigma} d \mu(x) d \mu(x') \int_0^\infty \frac{d t}{\hbar} \, \frac{t}{\hbar} e^{-\alpha_* t / \hbar} K_t(x,x') \right]^2 \,.  
\end{align}
We note that  the normalization integral in the denominator on the last line in Eq.~(\ref{zz}) can be reorganized as,
\begin{align}
& \iint_{\Sigma \times \Sigma} d \mu(x) d \mu(x') \int_0^\infty \frac{d t}{\hbar} \, \frac{t}{\hbar} e^{-\alpha_* t / \hbar} K_t(x,x') \nonumber \\
& \qquad = \frac{2}{\pi} \iint_{\Sigma \times \Sigma} d \mu(x) d \mu(x') \int_0^\infty \frac{d u}{\hbar} \, \sqrt{\frac{u}{\hbar}} \int_0^\infty \frac{d v}{\hbar} \, \sqrt{\frac{\hbar}{v}} e^{-\alpha_* (u+v) / \hbar} K_{u+v}(x,x') \,.
\end{align} 
It is easy to obtain the equality above by the following change of variables $t=u+v,t'=u-v$ in the double integral over the time variables on the right-hand side of the equation. The semi-group property is applied in order to decompose the heat kernel into a convolution of two heat kernels, thereby making the expression suitable for the application of the Cauchy-Schwarz inequality to the right-hand side of the expression above. 
\begin{align}
& \iint_{\Sigma \times \Sigma} d \mu(x) d \mu(x') \int_0^\infty \frac{d t}{\hbar} \, \frac{t}{\hbar} e^{-\alpha_* t / \hbar} K_t(x,x') \nonumber \\
& \qquad = \frac{2}{\pi} \int_{\cal M} d\tilde \mu(\tilde z) \iint_{\Sigma \times \Sigma} d \mu(x) d \mu(x') \int_0^\infty \frac{d u}{\hbar} \, \sqrt{\frac{u}{\hbar}}e^{-\alpha_*u/\hbar}K_u(x,\tilde z) \int_0^\infty \frac{d v}{\hbar} \, \sqrt{\frac{\hbar}{v}} e^{-\alpha_* v / \hbar} K_{v}(\tilde z,x') \,.
\end{align} 
Applying the Cauchy-Schwarz inequality and utilizing again  the semi-group property of the heat kernel in the resulting integrals  give
\begin{align}\label{t}
& \iint_{\Sigma \times \Sigma} d \mu(x) d \mu(x') \int_0^\infty \frac{d t}{\hbar} \, \frac{t}{\hbar} e^{-\alpha_* t / \hbar} K_t(x,x') \nonumber \\
& \qquad \leqslant \frac{2}{\pi} \left[ \iint_{\Sigma \times \Sigma} d \mu(x) d \mu(x') \int_0^\infty \frac{d u}{\hbar} \int_0^\infty \frac{d u'}{\hbar} \, \sqrt{\frac{u u'}{\hbar}} e^{-\alpha_* (u+u') / \hbar} K_{u+u'}(x,x') \right]^{1/2} \nonumber \\
& \qquad \times \left[ \iint_{\Sigma \times \Sigma} d \mu(y) d \mu(y') \int_0^\infty \frac{d v}{\hbar} \int_0^\infty \frac{d v'}{\hbar} \, \sqrt{\frac{\hbar^2}{v v'}} e^{-\alpha_* (v+v') / \hbar} K_{v+v'}(y,y') \right]^{1/2} \,.
\end{align}
Performing the following change of variables $t=u+u',t'=u-u'$ for the first and similarly defining  $s,s'$ for the other  double time integral on both lines on the right-hand side, and performing the integrals,  transforms the inequality~(\ref{t}) into
\begin{align}\label{tt}
& \sqrt{2} \iint_{\Sigma \times \Sigma} d \mu(x) d \mu(x') \int_0^\infty \frac{d t}{\hbar} \, \frac{t}{\hbar} e^{-\alpha_* t / \hbar} K_t(x,x') \nonumber \\
& \qquad \leqslant \left[ \iint_{\Sigma \times \Sigma} d \mu(x) d \mu(x') \int_0^\infty \frac{d t}{\hbar} \, \frac{t^2}{\hbar^2} e^{-\alpha_* t / \hbar} K_t(x,x') \right]^{1/2} \nonumber \\
& \qquad \times \left[ \iint_{\Sigma \times \Sigma} d \mu(y) d \mu(y') \int_0^\infty \frac{d s}{\hbar} \, e^{-\alpha_* s / \hbar} K_s(y,y') \right]^{1/2} \,.
\end{align}
If we replace the last line in Eq.~(\ref{zz}) by the inequality~(\ref{tt}), then we obtain
\begin{align}
\frac{\partial^2 E(\alpha)}{\partial \alpha^2} \Big \vert_{\alpha=\alpha_*} & \geqslant 0\,,
\end{align}
which infers that our extremum is actually a minimum. As said before we choose $-\alpha_*$ as the bound-state energy $E_b=-\nu_*^2$ so Eq.~(\ref{lam}) becomes
\begin{align}\label{lamm}
\frac{1}{\lambda} &= \frac{1}{V(\Sigma)} \iint_{\Sigma \time \Sigma} d \mu(x) d \mu(x') \int_0^\infty \frac{d t}{\hbar} \, e^{- \nu_*^2 t / \hbar} K_t(x,x') \,.
\end{align}
In case the ambient manifold is Euclidean space, due to the explicit form of the heat kernel one may explicitly execute the time integral and reach the special case in the limit $\nu_* \rightarrow 0$, which is first given in Proposition 6.1 in~\cite{ex5}. Eq.~(\ref{lamm}) suggests that this particular choice of the coupling constant for a given bound-state energy definitely allows us to construct the correct Hamiltonian of the model ab initio, thereby studying the bound-state structure thereof by the corresponding principal matrix. The on-diagonal part of the principal matrix for a collection of submanifolds, which admits bound states,  via Eq.(\ref{lamm}) takes the following form,
\begin{align}\label{phi}
\Phi_{ii}(-\nu^2) &= \frac{1}{V(\Sigma_i)} \iint_{\Sigma_i \times \Sigma_i} d \mu (x) d \mu (x') \int_0^\infty \frac{d t}{\hbar} \, \left( e^{-\nu_{i*}^2 t/ \hbar} - e^{-\nu^2 t / \hbar} \right) K_t(x,x') \,,
\end{align}
such that $\nu \in (0,\infty)$. Note that if Eq.(\ref{lamm}) has a solution $\nu_{i*}$ for the   given coupling constant $\lambda_i$, {\it we are allowed to replace the coupling constants by the corresponding bound states} as is done here. It is of course an interesting question to ask under what conditions there are solutions, which we will address subsequently. Another interesting direction, which will not be pursued here,  would be to study the cases for which the coupling is not strong enough to admit a bound state. Then scattering cross section will be an interesting problem.  The possibility of  appearance of bound states when many such surfaces put together will be another interesting direction to investigate,  as we will comment at the end of our work in a related context. 
The full Green's function stated in Eq.~(\ref{fg}), thus, has a well-defined limit, given by
\begin{align}
& R(\tilde x,\tilde x'\vert E) = R_0(\tilde x,\tilde x' \vert E) \nonumber\\
&\qquad \ + \Bigg[\frac{1}{\sqrt{V(\Sigma_i)}} \int_{\Sigma_i} d \mu(x)  R_0(\tilde x,x\vert E)\Bigg] \Phi_{ij}^{-1}(E)\Bigg[\frac{1}{\sqrt{V(\Sigma_j)}}\int_{\Sigma_j} d\mu(x')R_0(x',\tilde x' \vert E)\Bigg] \,, 
\end{align}
which at present  is defined only for $E<0$ along the real axis, but should be analytically continued to
 $E \in \mathbb{C}$. Although there are general results in this direction it is interesting to see this directly. Note that the free resolvent can be analytically continued so as to  have  a branch cut along the real axis  if the manifold is noncompact with the conditions as specified previously; otherwise it will have poles at the eigenvalues of the Laplacian along the real axis. 

Let us consider for simplicity a single surface, the analytic structure of  $\Phi(E)$ as a function of $E$ can be understood by using Theorem B in Section 2.12 in \cite{rosenblum}: Consider the kernel function,
\begin{align}
 F(\zeta, \xi)=\begin{cases} {\Phi(\zeta)-\Phi(\xi)\over \zeta-\xi} & \ \ \text{ if } \zeta\neq \xi \,, \\ 
{d\Phi\over d\zeta} & \ \  \text { if } \zeta=\xi \,.
\end{cases} 
 \end{align} 
Since $\Phi(E)$ is well-defined over a domain $\Delta=(-\infty, 0)$ along  the real axis, to test whether it has an analytic continuation $\tilde \Phi$ over the whole complex plane except possibly the positive real axis, satisfying  $\tilde \Phi|_\Delta=\Phi$, it is enough to show that for any complex valued function $f(\zeta)$ with compact support over $\Delta$ the kernel $F$ has the following positivity property;
\begin{align}
\int_\Delta \int_\Delta f(\zeta) F(\zeta, \xi) \bar f(\xi) \geq 0.
\end{align}
It is an instructive  exercise to check this explicitly in our case, a computation shows that the above quadratic form becomes;
\begin{align}
\int_{\cal M} d\tilde \mu(\tilde x) \Bigg| \int_\Delta d\zeta \, f(\zeta) \int_\Sigma d\mu(y)\; \int_0^\infty {dt\over \hbar}\,
K_t(y, \tilde x) e^{t\zeta/ \hbar}\Bigg|^2,
\end{align} 
which is strictly positive. Indeed the resulting analytic continuation $\tilde \Phi$ is a Nevallina function; this approach can be generalized to the matrix valued function $\Phi_{ij}(E)$ as well. We will denote this analytic continuation  by the same letter $\Phi$ to keep our notation simple.
Hence, the resolvent indeed is well defined over the complex plane, may have isolated poles at the zeros of the matrix $\Phi(E)$ along the negative real axis, and has a branch cut along the positive real axis for the noncompact ambient manifolds with asymptotically nonnegative Ricci tensor.

In the case of several submanifolds each of which is supporting a bound state, we may search for the bound state energies.  This means that we are looking for the solution of the following equation for $E_b=-\nu_*^2$, 
\begin{align}
\Phi_{ij}(E_b)A_j & = 0 \,,
\end{align} 
the consequences of which will be further discussed in Section~\ref{intermezzo}.

\section{The range of the coupling constant}\label{s3}

The exact relation between the interaction strength and the bound-state energy, defined by Eq.~(\ref{lamm}), also allows us to determine a lower bound on the possible values of the coupling constant  to assure that we find a preassigned value of the  bound-state energy. Moreover, we can use this approach to  assert that a bound-state is observed as soon as a  critical value of the coupling constant $\lambda_{C}$ is exceeded. 

In this respect, our main concern is now to lessen the right-hand side of Eq.~(\ref{lamm}) so as to find an estimate from above on  $\lambda_C$ while we are taking the limit $\nu_*\mapsto 0^+$. In doing so, the right-hand side is first made smaller via applying the Cheeger-Yau bound~\cite{chow}. This remarkable theorem  states that if $(\mathcal{M},g)$ is a complete Riemannian manifold with $Ric \geqslant (n-1) K$ for some $K \in \mathbb{R}$, then the heat kernel satisfies 
\begin{align}
K_t(x,x') &\geqslant K_t^{K} (d_g(x,x'))\,,  
\end{align}
where $K_t^{K} (d_g(x,x'))$ is the simply connected, complete, $n$-dimensional manifold of constant sectional curvature $K$. Henceforth, we will denote the metric distance on the ambient manifold by $d_\mathcal{M}$ and the one on the submanifold by $d_\Sigma$. As a corollary, given in Refs.~\cite{chow,dav}, if the manifold has a non-negative Ricci curvature, then the heat kernel of the ambient manifold $\mathcal{M}$ can be estimated from below by
\begin{align} \label{lkp}
K_t(\tilde x,\tilde x') &\geqslant \frac{m^{3/2} \exp \left( - \frac{m d_\mathcal{M}(\tilde x,\tilde x')}{2 \hbar t} \right)}{(2 \pi \hbar t)^{3/2}} \,,
\end{align}
which is written in terms of the physical quantities. 

Let us consider the non-negative Ricci case first in our problem. Recall that under this condition, the spectrum of the Laplacian is bounded from below by zero. There is a very natural estimate for the range of the coupling constant given by a purely geometric expression for a given value of $\nu_*$:
\begin{align}
\frac{1}{\lambda} \geqslant \frac{m}{2 \pi \hbar^2 V(\Sigma)} \iint_{\Sigma \times \Sigma} d \mu(x) d \mu(x') \, \frac{\exp \left( - \sqrt{\frac{2 m}{\hbar^2}} \nu_* d_{\mathcal{M}}(x,x') \right)}{d_{\mathcal{M}}(x,x')} \,.  
\end{align}
In general this integral on the right-hand side is hard to estimate, and we will elaborate on this point below. One possibility is to use the diameter of the embedded submanifold since $d_{\mathcal{M}}(x,x') \leqslant \max_{x,x' \in \Sigma} d_{\mathcal{M}}(x,x') = d_{\mathcal{M}}(\Sigma)$. Here $d_{\mathcal{M}}(\Sigma)$ is the diameter of the submanifold which is measured in the ambient manifold. Thus, the inverse coupling constant satisfies
\begin{align}
\frac{1}{\lambda} \geqslant \frac{m V(\Sigma)}{2 \pi \hbar^2} \frac{\exp \left( - \sqrt{\frac{2 m}{\hbar^2}} \nu_* d_{\mathcal{M}}(\Sigma) \right)}{d_{\mathcal{M}}(\Sigma)} \,,
\end{align}
where $V(\Sigma)$ is the surface area of the embedded submanifold.
This inequality demonstrates that there is always a critical coupling constant beyond which there will always be a bound state arbitrarily small binding energy in the limit $\nu_* \rightarrow 0^+$.
\begin{align}
\frac{1}{\lambda_C} \geqslant \frac{m V(\Sigma)}{2 \pi \hbar^2} \frac{1}{d_{\mathcal{M}}(\Sigma)} \,.  
\end{align}
This estimate of the critical coupling can be a good estimate for submanifolds which are ``tightly'' embedded, that is to say $V_\mathcal{M}(\Sigma) / d_{\mathcal{M}}^3(\Sigma) \sim O(1)$, where $V_\mathcal{M}(\Sigma)$ is the volume of the region enclosed by the submanifold. This crude idea is to be investigated for various distinct geometries.
 
Incidentally, the existence of critical coupling should be contrasted with the case of curves embedded into two dimensional ambient manifolds \cite{bt1}; using the notation in the original work, in that case the equation for the critical coupling becomes (for positive Ricci case for simplicity):
\begin{eqnarray}
    {1\over \lambda_C}&\geqslant& {m\over 2L \pi \hbar^2}\int_0^\infty  \int_{\Gamma\times \Gamma}\,ds\, ds' \exp\Big(-{md_{\cal M}^2(\gamma(s),\gamma(s'))\over 2\hbar t}\Big){dt\over t}\nonumber\\
&\geqslant& {m L\over 2 \pi \hbar^2}\int_0^\infty   \exp\Big(-{md_{\cal M}^2(\Gamma)\over 2\hbar t}\Big){dt\over t}
.\end{eqnarray}
In the last line we used the lower  estimate of replacing the geodesic distance by the diameter $d_{\cal M} (\Gamma)$ of the curve
$\Gamma$, one sees that  this integral diverges as $t\to \infty$. Thereby showing that an arbitrarily weak coupling will lead to a  bound state. From this point of view too,  the study of surfaces thus is much richer.

 The distance between the points $x,x'\in\Sigma\subset\mathcal{M}$ is greater when measured with respect to the induced metric on the submanifold, thereby replacing $d_\mathcal{M}$ by $d_\Sigma$ in the exponent obviously makes the integral smaller. Inserting the inequality~(\ref{lkp}) into Eq.(\ref{lamm}) gives us the following inequality,
\begin{align}
\frac{1}{\lambda} \geqslant \frac{1}{V(\Sigma)} \iint_{\Sigma \times \Sigma} d \mu(x)  d \mu(x') \int_0^\infty \frac{d t}{\hbar} \, e^{-\nu_*^2 t / \hbar} \frac{m^{3/2} \exp \left( - \frac{m d_\Sigma(x,x')}{2 \hbar t} \right)}{(2 \pi \hbar t)^{3/2}} \,.
\end{align}
The $t$-integral here is an elementary one, and our inequality by calculating this integral becomes
\begin{align}\label{l1}
\frac{1}{\lambda} \geqslant \frac{m}{2 \pi \hbar^2 V(\Sigma)} \iint_{\Sigma \times \Sigma} d \mu(x) d \mu(x') \, \frac{\exp \left( - \sqrt{\frac{2 m}{\hbar^2}} \nu_* d_\Sigma(x,x') \right)}{d_\Sigma(x,x')} \,.  
\end{align}
In general, this replacement could be a crude estimate. Yet, in certain geometries, for example large cigar-like protrusions coming off the surface, the geodesic distance in the ambient manifold and the one along the submanifold may become comparable over large areas.\footnote{We would like to thank the referee for pointing out this example.} The necessity of examining special geometries rests upon the following observations: The geodesic distance between two points on the surface varies in a very complicated way for a general ambient manifold. Moreover, the intrinsic geometry of the embedded surface depends on the genus of the compact submanifold as well, whereby understanding such embeddings in general is already a hard problem. Therefore, we plan to investigate those possibilities for special geometries in further studies.
In order to proceed further, we need to equip ourselves with crucial geometric information regarding the submanifold $\Sigma$, such as how its volume grows. The so-called volume comparison theorems~\cite{chow,gall} and the usage thereof are of great importance at this point. Since we only consider the submanifolds, whose sectional curvatures satisfy $L \leqslant Sec \leqslant H$ such that $L,H \in \mathbb{R}$, these theorems become applicable to our problem. Since we aim for some simple explicit estimates on the critical coupling constant, we make the following further assumption on the submanifold such that the lower bound on the sectional curvature of the submanifold is chosen to be $L=(1-\varepsilon) H$ provided $\varepsilon \in \mathbb{R}$ is a small number. Note that $L$ stands for $(1-\varepsilon) H$ throughout this section.

The volume comparison theorems  provide a relation between the volume of the manifold under consideration and the volume of the model space, defined by the bound either on the sectional or on the Ricci curvature of the original manifold. Since 2-dimensional submanifolds only involve in our problem, the lower bound on the Ricci curvature and the one on the sectional curvature can be used interchangeably.

We can choose an arbitrary point on $\Sigma$, and consider the geodesic ball centered at this point.  The exponential map at this point provides geodesic polar coordinates. Therefore, a geodesic ball $B_p(r)$ is defined such that $B_p(r)=\{ x=\exp_p(s\theta) \vert \theta \in S^{n-1}, 0 \leqslant s \leqslant r \}$, here   $\theta \in S^{n-1}\subset \Sigma_p$ denotes an element in the unit sphere of $\Sigma_p$, the tangent space at  point $p$ in $\Sigma$. The Riemannian volume of this ball is denoted by $V(B_p(r))$. As $\Sigma$ is compact and satisfies a lower bound $Sec \geqslant L$, every geodesic of length $\geqslant \pi L^{-1/2} $ has conjugate points, whereupon the diameter of $\Sigma$ satisfies $d(\Sigma) \leqslant \pi L^{-1/2}$ due to Bonnet-Myers theorem \cite{chee}. We will, however, restrict ourselves to a convex neighborhood since our focus in this section is on achieving a lower bound. 
Note that  the integrand in the inequality~(\ref{l1}) is only the function of the geodesic distance between the points $x$ and $x'$ on the submanifold, we choose one of the points and consider the geodesic ball centered at that point. The geodesic distance between these points turns into the radial distance connecting them. Therefore, the double integral becomes an ordered one.

It is time to employ the volume comparison theorem due to Bishop-G\"{u}nter \cite{gall}. It states that, for a complete Riemannian manifold $(\Sigma,g)$ with $Sec \leqslant H$ such that $H \in \mathbb{R}$, as long as  $B_p(r)$ does not  meet the cut-locus of $p$,  the ratio
\begin{align}
r &\rightarrow \frac{V(B_p(r))}{V(B_p^{H}(r))} \,,
\end{align} 
is a non-decreasing function, which tends to 1 as $r$ goes to 0. Here, $V(B_p^{H}(r))$ denotes the Riemannian volume of a ball of radius $r$ in the complete simply connected Riemannian manifold with constant curvature $H$, which is called a space form of constant curvature $H$. In particular we have the following infinitesimal volume comparison formula,
\begin{align}
d \mu_\Sigma &\geqslant d \mu_{H} \nonumber \\
J(r,\theta) \, d r d \theta & \geqslant J_{H}(r) \, d r d \theta \,,
\end{align}
where $J(r,\theta)$ is the Jacobian of the exponential map, that is the volume density in geodesic polar coordinates. For a space form of constant curvature $K$, this Jacobian has particular forms, depending on $K$, and they are given by~\cite{chow},
\begin{align}\label{j}
J_K (r) = \left\{ \begin{array}{rl}
\frac{1}{\sqrt{K}} \sin (\sqrt{K}r) & \text{if } K > 0\,, \\ 
r & \text{if } K = 0 \,, \\ 
\frac{1}{\sqrt{K}} \sinh (\sqrt{K}r) & \text{if } -K < 0\,. \\
\end{array} \right.
\end{align}
Note that the above formulae for the Jacobian of the exponential map $J_K(r)$ hold only if $r\in(0,\pi/\sqrt{K})$ is assumed when $K>0$, and for all $r>0$ when $-K \leq 0$. After fixing one of the points, the inequality~(\ref{l1}) takes the following form in the geodesic polar coordinates,
\begin{align}
\frac{1}{\lambda} &\geqslant \frac{m}{2\pi\hbar^2 V(\Sigma)} \int_\Sigma d \mu  \int_0^{2 \pi} d \theta \int_0^{\rho(\theta)} d r \, \frac{J(r,\theta)}{r} \exp \left( - \sqrt{\frac{2 m}{\hbar^2}} \nu_* r \right)\,.
\end{align}
Now, we replace $\rho(\theta)$ by $\rho_*=\min_\theta \rho(\theta)$ such that $\rho_* \leqslant conv.rad(\Sigma)$, where $conv.rad(\Sigma)$ denotes the convexity radius of $\Sigma$ \cite{petersen}. Applying the volume comparison theorem thereafter leads to
\begin{align}\label{l2}
\frac{1}{\lambda} &\geqslant \frac{m}{\hbar^2}\int_0^{\rho_*} d r \, \frac{J_{H}(r)}{r} \exp \left( - \sqrt{\frac{2 m}{\hbar^2}} \nu_* r \right)\,.
\end{align}
Let us analyze three cases that are distinguished by their sectional curvatures:

{\large \textbf{Case $(H=0)$}} The integral need not be estimated due to its simplicity. The following result by computing the integral is  obtained,
\begin{align}
\frac{1}{\lambda} & \geqslant \frac{m}{\hbar^2} \frac{1-\exp \left( - \sqrt{\frac{2m}{\hbar^2}} \nu_* \rho_* \right)}{\sqrt{\frac{2m}{\hbar^2}} \nu_*} \,.
\end{align}
If one takes the limit $\nu_* \rightarrow 0$, then an upper bound on the  critical value of the interaction strength, namely $\lambda_C$, is obtained, and it is given by 
\begin{align}
\frac{1}{\lambda_C} & \geqslant \frac{m}{\hbar^2} \rho_* \,.
\end{align}
As long as $\lambda$ is chosen greater than the critical value $\lambda_C$, we assure the existence of a bound state solution, and {\it the bound we obtained, therefore,  provides a lower bound for the allowed range of coupling constants which leads to a bound state}.

{\large \textbf{Case $(H > 0)$}}In this case we have the following inequality,
\begin{align}
\frac{1}{\lambda} &\geqslant \frac{m}{\hbar^2}\int_0^{\rho_*} d r \, \frac{\sin(\sqrt{H} r)}{\sqrt{H} r} \exp \left( - \sqrt{\frac{2 m}{\hbar^2}} \nu_* r \right)\,.
\end{align}
We can make the right-hand side smaller via estimating $\sin(\xi)/\xi$ by $(1+\cos \xi)/2$ using the Hermite-Hadamard inequality \cite{mit}. With this replacement we obtain
\begin{align}
\frac{1}{\lambda} &\geqslant \frac{m}{\sqrt{H}\hbar^2}\int_0^{\rho_* \sqrt{H}} d \xi \, \frac{1+\cos \xi}{2} \exp \left( - \sqrt{\frac{2 m}{H \hbar^2}} \nu_* \xi \right)\,.
\end{align} 
The integral can be estimated further by the integral Chebyshev inequality \cite{mit}. It states that
\begin{align}
\frac{1}{b-a} \int_a^b d \xi \, f(\xi) g(\xi) & \geqslant \left[ \frac{1}{b-a} \int_a^b d \xi \, f(\xi) \right] \left[ \frac{1}{b-a} \int_a^b d \xi \, g(\xi) \right] \,,
\end{align}
if $f$ and $g$ are two monotonic functions of the same monotonicity type. Applying this estimate to our inequality leads to
\begin{align}
\frac{1}{\lambda} & \geqslant \frac{m}{2 \sqrt{H} \hbar^2}  \frac{1-\exp \left( - \sqrt{\frac{2 m}{\hbar^2}} \nu_* \rho_* \right)}{\sqrt{\frac{2m}{\hbar^2}} \nu_* \rho_*} \left[ \sqrt{H} \rho_* + \sin (\sqrt{H} \rho_*) \right] \,.
\end{align} 
The limit $\nu_* \rightarrow 0$ allows us to obtain $\lambda_C$, and it follows that
\begin{align}
\frac{1}{\lambda_C} \geqslant \frac{m}{2 \sqrt{H} \hbar^2} \left[ \sqrt{H} \rho_* + \sin (\sqrt{H} \rho_*) \right] \,.
\end{align}

{\large \textbf{Remark.}} Although this inequality and the other ones in this section assures the occurrence of a bound state when the coupling constant is chosen to be greater a critical value, they are insensitive to the change in the sectional curvature of the submanifold due to a slight deformation thereof, for example  a deformation in the normal direction to the submanifold. The volume comparison theorem, which we are using to obtain the upper bound on the coupling constant, enforces us to choose the region of which the sectional curvature is the highest, while integrating  on the submanifold,  so as to reduce the estimated area. However, if we infinitesimally deform the submanifold in the normal direction, then we can intuitively observe that the sectional curvature of the corresponding space form, which is a sphere in this case, will become smaller due to the increase of its radius in the ambient manifold. This suggest that we should use a volume comparison theorem in the opposite direction that will enlarge its area by a small amount, thereby obtaining a lower bound for the critical coupling constant. This allows us to choose the smallest value of the sectional curvature while moving on the submanifold. We can intuitively see how this argument works in a simple example. Let us consider a sphere in space. The eigenvectors of its shape operator are the principal directions, and their eigenvalues are the principal curvatures. They are equal to each other, and the values thereof equal the inverse of the radius of the sphere. If we slightly deform this sphere in one of the principal direction while preserving its total area, then we obtain a prolate ellipsoid. It is obvious that the regions of the submanifold of smaller curvature, which, in other words, have larger area, contributes more to the integral, which determines the inverse of the coupling constant. The volume comparison theorem to be used for this purpose is the well-known Bishop-Gromov volume comparison theorem. Since we mainly use this theorem in the next section, the reader is invited to the next section for the details of this theorem. Let us choose the ambient manifold an Euclidean three space for simplicity. Then, we have the corresponding upper bound on the inverse critical coupling constant in the limit $\nu_* \rightarrow 0$:
\begin{align}
\frac{1}{\lambda_C} \leqslant \frac{m}{\hbar^2 \sqrt{1- \delta \kappa_g^*}} \int_0^{\rho^* \sqrt{L}} d \xi \, \frac{\sin\xi}{\xi} \,,
\end{align} 
where the geodesic distance measured in the ambient manifold is replaced by the one on the submanifold multiplied by $\sqrt{1- \delta \kappa^*}$, and $\rho^*=\max_\theta \rho(\theta)$. This replacement of the geodesic distance is mainly used in the next section, therefore, we would like to invite the reader to the next section for the details of this idea, as well. Now, the right-hand side is enlarged by estimating $\sin \xi / \xi$ by $\cos (\xi/2)$. Evaluating the integral gives the following lower bound on the critical coupling constant:
\begin{align}
\lambda_C \geqslant \frac{\hbar^2}{2m} \frac{\sqrt{(1- \delta \kappa_g^*) L}}{\sin\frac{\sqrt{L} \rho^*}{2}} \,.
\end{align}
The infinitesimal deformation under consideration makes the sectional curvature smaller: $L=(1-(1+a)\varepsilon) H$ where $a$ is a real number related to the deformation. As a result, the bound on the critical coupling constant diminishes under such deformation of the submanifold since the right-hand side of the inequality is a monotonically increasing function of the sectional curvature. This result is in accordance with the earlier result obtained in~\cite{ex5}, in which the authors show that the critical coupling constant decreases under a small enough area-preserving smooth radial deformation of the sphere. The above bound is still a crude estimate used only to understand the general behavior under deformations. The arguments can be generalized to other cases but for the sake of brevity we will refrain from this.  

{\large \textbf{Case $(-H<0)$}} The inequality this time becomes
\begin{align}
\frac{1}{\lambda} &\geqslant \frac{m}{\hbar^2}\int_0^{\rho_*} d r \, \frac{\sinh(\sqrt{H} r)}{\sqrt{H} r} \exp \left( - \sqrt{\frac{2 m}{\hbar^2}} \nu_* r \right)\,.
\end{align}
Approximating $\sinh \xi / \xi$ from below by $\cosh(\xi/2)$ via the Hermite-Hadamard inequality, and calculating the integral thereafter turn the inequality into
\begin{align}
\frac{1}{\lambda} &\geqslant \frac{2 m}{ \sqrt{H}\hbar^2} \frac{2 \sqrt{\frac{2 m}{H \hbar^2}} \nu_* \left[ -1 + \cosh \left( \frac{\sqrt{H} \rho_*}{2} \right) e^{ - \sqrt{\frac{2 m}{\hbar^2}} \nu_* \rho_*} \right] + \sinh \left( \frac{\sqrt{H} \rho_*}{2} \right) e^{ - \sqrt{\frac{2 m}{\hbar^2}} \nu_* \rho_*}}{1-\frac{8 m \nu_*^2}{H \hbar^2}} \,.
\end{align}
The desired bound on $\lambda_C$ follows from the limit $\nu_* \rightarrow 0$,
\begin{align}
\frac{1}{\lambda_C} &\geqslant \frac{2m}{\sqrt{H} \hbar^2} \sinh \left( \frac{\sqrt{H} \rho_*}{2} \right) \,.
\end{align}

We would like to remind the reader that, so far we have considered the case in which the ambient manifold $\mathcal{M}$ has a non-negative Ricci curvature. We can generalize further, and repeat the calculations for $\mathcal{M}$ with  Ricci curvature bounded from below by a negative constant. This time,  the heat kernel of the model space with constant negative sectional curvature is needed and fortunately, it is exactly known~\cite{dav}, 
\begin{align}
K_t(x,x') &= \frac{m^{3/2}}{(2 \pi \hbar t)^{3/2}} \frac{\sqrt{K} d_\mathcal{M}(x,x')}{\sinh (\sqrt{K} d_\mathcal{M}(x,x'))} \exp \left[ - \frac{K \hbar t}{2 m} - \frac{m d_\mathcal{M}^2(x,x')}{2 \hbar t}\right] \,.
\end{align}
The analogous result to the inequality~(\ref{l2}) via repeating the previous steps is found as
\begin{align}\label{l3}
\frac{1}{\lambda} &\geqslant \frac{m \sqrt{\pi}}{\hbar^2} \int_0^{\rho_*} d r \, \frac{ \sqrt{K} J_{H} (r)}{\sinh(\sqrt{K} r)} \exp \left( - \sqrt{\frac{2 m \nu_*^2 + K \hbar^2}{\pi \hbar^2}} r \right)\,. 
\end{align}
Let us look at the possible cases as having done before:

{\large\textbf{Case $(H = 0)$}} Due to vanishing sectional curvature condition, we have $\xi / \sinh \xi$ in the integrand, and replacing it by $e^{-\xi}$ makes the right-hand side smaller. After calculating the integral, we have
\begin{align}
\frac{1}{\lambda} &\geqslant \frac{m \pi \sqrt{K}}{\hbar} \frac{1 - \exp \left( - \sqrt{K} \rho_* - \sqrt{\frac{2 m \nu_*^2 + K \hbar^2}{\pi \hbar^2}} \rho_* \right)}{\sqrt{K \pi} \hbar + \sqrt{2 m \nu_*^2 + K \hbar^2}} \,.
\end{align}
The corresponding critical value of the coupling constant after taking the  limit is equal to
\begin{align}
\frac{1}{\lambda_C} &\geqslant \frac{m \pi}{\hbar^2 (1 + \sqrt{\pi})} \exp \left(- \frac{\sqrt{K} (1+\sqrt{\pi}) \rho_*}{\sqrt{\pi}} \right) \,.
\end{align}

{\large \textbf{Case $(H>0)$}}Inequality~(\ref{l3}) becomes complicated since both trigonometric and hyperbolic functions involve in it, which turns the inequality into
\begin{align}
\frac{1}{\lambda} &\geqslant \frac{m \sqrt{\pi}}{\hbar^2} \int_0^{\rho_*} d r \, \frac{\sqrt{K} \sin (\sqrt{H} r)}{\sqrt{H} \sinh (\sqrt{K}r)} \exp \left( - \sqrt{\frac{2 m \nu_*^2 + K \hbar^2}{\pi \hbar^2}} r \right)\,.
\end{align}  
We can, however, simplify these terms by the estimates used in the previous cases, so one has
\begin{align}
\frac{1}{\lambda} &\geqslant \frac{m \sqrt{\pi}}{\hbar^2} \int_0^{\rho_*} d r \, \frac{1+\cos (\sqrt{H} r)}{2} \exp \left(-\sqrt{K}r - \sqrt{\frac{2 m \nu_*^2 + K \hbar^2}{\pi \hbar^2}} r \right)\,.
\end{align}
As having done before, we will employ the integral Chebyshev inequality in the right-hand side of the expression above, which results in
\begin{align}
\frac{1}{\lambda} &\geqslant \frac{m \pi}{2 \hbar} \frac{1-\exp\left(- \sqrt{K} \rho_* - \sqrt{\frac{2 m \nu_*^2 + K \hbar^2}{\pi}} \rho_* \right)}{\sqrt{K \pi}  \hbar + \sqrt{2 m \nu_*^2 + K \hbar^2}} \left[ \rho_* + \frac{\sin(\sqrt{H} \rho_*)}{\sqrt{H}} \right] \,.
\end{align}  
One can take the $\nu_* \rightarrow 0$ limit, which gives us the following bound for $\lambda_C$,
\begin{align}
\frac{1}{\lambda_C} &\geqslant \frac{m \pi}{2 \sqrt{H K} \hbar^2} \frac{1 - \exp \left[ - \sqrt{K} \left(1 + \frac{1}{\sqrt{\pi}} \right) \rho_* \right]}{1 + \sqrt{\pi}} \left[ \sqrt{H} \rho_* + \sin (\sqrt{H} \rho_*) \right] \,.
\end{align}

{\large \textbf{Case $(-H<0)$}} Although inequality~(\ref{l3}) at first sight, does not  look as  complicated as the previous one, still it is. It contains a ratio of hyperbolic functions besides an exponential after plugging $J_{H}$ in it. Therefore, it follows
\begin{align}
\frac{1}{\lambda} &\geqslant \frac{m \sqrt{\pi}}{\hbar^2} \int_0^{\rho_*} d r \, \frac{ \sqrt{K} \sinh (\sqrt{H} r)}{\sqrt{H} \sinh(\sqrt{K} r)} \exp \left( - \sqrt{\frac{2 m \nu_*^2 + K \hbar^2}{\pi \hbar^2}} r \right)\,.
\end{align}
It is now assumed that the condition $H < K$ is satisfied whereby one can estimate the aforesaid ratio from below by $\exp (- \sqrt{H} r - \sqrt{K} r) \sqrt{H} / \sqrt{K}$. The inequality~(\ref{l3}) and its $\nu_* \rightarrow 0$ limit are, respectively, given by
\begin{align}
\frac{1}{\lambda} &\geqslant \frac{m \sqrt{\pi}}{\hbar} \frac{1 - \exp \left[ - (\sqrt{H} + \sqrt{K}) \rho_* - \sqrt{\frac{2m \nu_*^2 + K \hbar^2}{\pi}} \rho_* \right]}{(\sqrt{H} + \sqrt{K}) \hbar + \sqrt{\frac{2 m \nu_*^2 + K \hbar^2}{\pi}}} \,, \\
\frac{1}{\lambda_C} &\geqslant \frac{m \sqrt{\pi}}{\hbar^2} \frac{1-\exp \left[ - \sqrt{H} \rho_* - \sqrt{K} (1 + \frac{1}{\sqrt{\pi}}) \rho_* \right]}{\sqrt{H} + \sqrt{K} (1 + \frac{1}{\sqrt{\pi}})} \,.
\end{align}

This concludes our discussion on the possible ranges of the coupling constants so as to assure the existence of a bound state. We will next verify that the theory is indeed finite.

\section{Finiteness of the principal matrix}\label{s4}

In this section we aim to show that the model does not need any renormalization. For this purpose, we need to bring the off-diagonal upper bound of the heat kernel into play. In Ref.~\cite{et} this bound of the heat kernel on a compact manifold is shown to be given by
\begin{align}\label{ch}
K_t(x,x') &\leqslant \left[ \frac{C_1}{V(\mathcal{M})} + C_2 \left( \frac{m}{2 \pi \hbar t} \right)^{3/2} \exp \left( - \frac{m d_\mathcal{M}^2(x,x')}{2 C_3 \hbar t} \right) \right] \,,
\end{align}
where $C_1,C_2$ and $C_3$ are some absolute constants whose values particularly depend on the geometric features of the manifold in question. If the manifold, however, is a noncompact one, then the term weighted with the volume of the manifold does not exist. That is why, we will reach the result just for the former kind.   

The first step  is to insert our heat kernel estimate into the quantity desired to be analyzed. For this task, the replacement on-diagonal part of the principal operator given by Eq.~(\ref{phi}) suffices since the off-diagonal elements have integrals over non-overlapping submanifolds. Thus,    with $\nu$ chosen larger than $\nu_{i*}$,  we have,
\begin{align}\label{phik}
\Phi_{ii}(-\nu^2) &\leqslant \frac{1}{V(\Sigma_i)} \iint_{\Sigma_i \times \Sigma_i} d \mu (x) d \mu (x') \int_0^\infty \frac{d t}{\hbar} \left( e^{-\nu_{i*}^2 t/ \hbar} - e^{-\nu^2 t / \hbar} \right) \nonumber \\
& \qquad \times \left[ \frac{C_1}{V(\mathcal{M})} + C_2 \left( \frac{m}{2 \pi \hbar t} \right)^{3/2} \exp \left( - \frac{m d_\mathcal{M}^2(x,x')}{2 C_3 \hbar t} \right) \right] \,.
\end{align}
Since this inequality comprises  two independent complicated expressions, we choose to analyze them separately. Let us look at the one with the volume term first, and denote it by $\mathtt{I}$. Integration over the variable $t$ leads to
\begin{align}\label{i1}
\mathtt{I} &= \frac{C_1}{V(\Sigma_i) V(\mathcal{M})} \sqrt{\frac{2 m}{C_3 \hbar^2}} \iint_{\Sigma_i \times \Sigma_i} d \mu (x) d \mu (x') \, \frac{d_\mathcal{M}(x,x')}{\nu_{i*}\nu} \nonumber \\
& \qquad \times \left[ \nu K_1 \left( \sqrt{\frac{2 m}{C_3 \hbar^2}} \nu_{i*} d_\mathcal{M}(x,x') \right) - \nu_{i*} K_1 \left( \sqrt{\frac{2 m}{C_3 \hbar^2}} \nu d_\mathcal{M}(x,x')\right) \right] \,, 
\end{align}    
$K_1(z)$ being the modified Bessel function of the second kind of order 1~\cite{ol}. Although the inequality $d_\mathcal{M}(x,x') < d_{\Sigma_i}(x,x') $ can be applied to the numerator following the measures, same replacement can not be done for the Bessel functions, because they are  monotonically decreasing functions.

In Ref.~\cite{bt1}, the authors consider a curve which is isometrically embedded in a Riemannian manifold. In any  sufficiently small neighborhood,  they obtain an inequality that relates the distance between two of its points in the ambient space, to the arclength of this small  segment of the  curve and  its curvature. There, it is  shown that
\begin{align}
\sqrt{1-\delta \kappa_g^*} \xi &\leqslant d_\mathcal{M}\left(\gamma(s),\gamma(s') \right) \,, \text{ with } 0<\delta< \frac{\kappa_g^*}{2} \,,
\end{align}
where $\gamma(s)$ is an arclength parametrized curve, and $\xi$ denotes the length of a segment thereof, defined by $\xi= \vert s - s'\vert$. Furthermore, $\delta$ is the maximum value that $\xi$ can attain in a geodesic ball centered at the point $s$, and $\kappa_g^*$ is the maximum value of the geodesic curvature along the curve. In our case, $\xi$ can be substituted for $d_{\Sigma_i}(x,x')$, so for the radial geodesic eventually, provided $\kappa_g$ should be replaced by the extrinsic curvature of this curve. It is obvious that a curve, which is a geodesic of a submanifold, can not possess a geodesic curvature. As a result, its curvature merely originates from the embedding of this submanifold into an ambient manifold whereby the value of its curvature is only determined by the second fundamental form of this embedding~\cite{lee}. We will not only apply this inequality, but also apply the inequality $K_1(z)<e^{-z}(1+1/z)$, which is shown in Ref.~\cite{bt2}, to our expression. As a consequence, Eq.~(\ref{i1}) takes the following form,
\begin{align}
\mathtt{I} &\leqslant \frac{C_1}{V(\Sigma_i) V(\mathcal{M})} \sqrt{\frac{2 m}{C_3 \hbar^2}} \iint_{\Sigma_i \times \Sigma_i} d \mu (x) d \mu (x') \nonumber \\ 
& \qquad \times \left\{ \left[d_{\Sigma_i}(x,x') + \frac{\sqrt{C_3 \hbar^2}}{\sqrt{2 m (1- \delta_i \kappa_i^*)} \nu_{i*}} \right] \frac{e^{-\sqrt{\frac{2 m (1- \delta_i \kappa_i^*)}{C_3 \hbar^2}} \nu_{i*} d_{\Sigma_i}(x,x')}}{\nu_{i*}} \right. \nonumber \\
& \qquad \qquad \left. - \left[ d_{\Sigma_i}(x,x') + \frac{\sqrt{C_3 \hbar^2}}{\sqrt{2 m (1- \delta_i \kappa_i^*)} \nu} \right] \frac{e^{-\sqrt{\frac{2 m (1- \delta_i \kappa_i^*)}{C_3 \hbar^2}} \nu d_{\Sigma_i}(x,x')}}{\nu} \right\} \,. 
\end{align}
The same geometric argument offered to estimate the integral in the previous section is used as a subsequent step. Thus, we put Eq.~(\ref{i1}) into a convenient form so as to be estimated from above by means of a volume comparison theorem which is different from the one used in the previous section. Now we have
\begin{align}
\mathtt{I} &\leqslant \frac{C_1}{V(\Sigma_i) V(\mathcal{M})} \sqrt{\frac{2 m}{C_3 \hbar^2}} \int_{\Sigma_i} d \mu \int_0^{2 \pi} d \theta \int_0^{\rho_i(\theta)} d r \, J(r,\theta) \nonumber \\
& \qquad \times \left\{ \left[r + \frac{\sqrt{C_3 \hbar^2}}{\sqrt{2 m (1- \delta_i \kappa_i^*)}\nu_{i*}} \right] \frac{e^{-\sqrt{\frac{2 m (1- \delta_i \kappa_i^*)}{C_3 \hbar^2}} \nu_{i*} r}}{\nu_{i*}} \right. \nonumber \\
& \qquad \qquad \left. - \left[r + \frac{\sqrt{C_3 \hbar^2}}{\sqrt{2 m (1- \delta_i \kappa_i^*)}\nu} \right] \frac{e^{-\sqrt{\frac{2 m (1- \delta_i \kappa_i^*)}{C_3 \hbar^2}} \nu r}}{\nu} \right\} \,.
\end{align}   

The required theorem is here known as the Bishop-Gromov volume comparison theorem~\cite{gall}. It states that, for a complete Riemannian manifold $(\Sigma,g)$ with $Ric \geqslant (n-1) L$, for  $L \in \mathbb{R}$,  as long as $B_p(r)$ does not meet the cut-locus of $p$, the ratio
\begin{align}
r &\rightarrow \frac{V(B_p(r))}{V(B_p^{L}(r))} \,,
\end{align} 
is a non-increasing function, which tends to 1 as $r$ goes to 0. $V(B_p^{L}(r))$ here denotes the Riemannian volume of a ball of radius $r$ in the space form of constant curvature $L$. In particular, the corresponding infinitesimal volume comparison formula is given by
\begin{align}
d \mu_\Sigma &\leqslant d \mu_L \,.
\end{align}
Now recall that the formula for the Jacobian of the exponential map holds for all $r>0$ when $-L \leq 0$ whereas when $L>0$,  it holds only if $r\in(0,\pi/\sqrt{L})$ is assumed. Recall also that we can employ the lower bound on the sectional curvature in place of the lower bound on the Ricci curvature due to the fact that the submanifolds considered in this model are two dimensional. We will substitute $\rho^* = \max_\theta \rho(\theta)$ for $\rho(\theta)$, and will apply the volume comparison theorem thereafter, giving us an upper bound to Eq.~(\ref{i1}) which becomes explicitly dependent  to the curvature bound $L$. It  follows that 
\begin{align}\label{i12}
\mathtt{I} &\leqslant \frac{2 \pi C_1}{V(\mathcal{M})} \sqrt{\frac{2 m}{C_3 \hbar^2}} \int_0^{\rho_i^*} d r \, J_{L_i}(r) \left\{ \left[r + \frac{\sqrt{C_3 \hbar^2}}{\sqrt{2 m (1- \delta \kappa^*)} \nu_{i*}}  \right] \frac{e^{-\sqrt{\frac{2 m (1- \delta \kappa^*)}{C_3 \hbar^2}} \nu_{i*} r}}{\nu_{i*}} \right. \nonumber \\ 
& \qquad \left. - \left[r + \frac{\sqrt{C_3 \hbar^2}}{\sqrt{2 m (1- \delta \kappa^*)} \nu}  \right] \frac{e^{-\sqrt{\frac{2 m (1- \delta \kappa^*)}{C_3 \hbar^2}} \nu r}}{\nu} \right\} \,.
\end{align}

Let us choose $-L_i<0$ and show that the expression~(\ref{i12}) can be made smaller than an arbitrarily large but finite quantity. We now have
\begin{align}
\mathtt{I} &\leqslant \frac{2 \pi C_1}{V(\mathcal{M})} \sqrt{\frac{2 m}{C_3 \hbar^2}} \int_0^{\rho_i^*} d r \, \frac{\sinh (\sqrt{L_i}r)}{\sqrt{L_i}} \left\{ \left[r + \frac{\sqrt{C_3 \hbar^2}}{\sqrt{2 m (1- \delta_i \kappa_i^*)} \nu_{i*}}  \right] \frac{e^{-\sqrt{\frac{2 m (1- \delta_i \kappa_i^*)}{C_3 \hbar^2}} \nu_{i*} r}}{\nu_{i*}} \right. \nonumber \\ 
& \qquad \left. - \left[r + \frac{\sqrt{C_3 \hbar^2}}{\sqrt{2 m (1- \delta_i \kappa_i^*)} \nu}  \right] \frac{e^{-\sqrt{\frac{2 m (1- \delta_i \kappa_i^*)}{C_3 \hbar^2}} \nu r}}{\nu} \right\} \,.
\end{align}
Since $\sinh (\sqrt{L_i} r)$ is a monotonically increasing function, taking it out of the integral by its value at $\rho_i^*$ makes the integral larger, and calculating the integral thereafter gives us
\begin{align}
\mathtt{I} &\leqslant \frac{2 \pi C_1}{V(\mathcal{M})} \sqrt{\frac{C_3 \hbar^2}{2 m}} \frac{\sinh (\sqrt{L_i}\rho_i^*)}{\sqrt{L_i}(1-\delta_i\kappa_i^*)} \left[ \frac{2 \left(1-e^{- \alpha_i \nu_{i*} \rho_i^*} \right) - \alpha_i \nu_{i*} \rho_i^* e^{- \alpha_i \nu_{i*} \rho_i^*}}{\nu_{i*}^3} \right. \nonumber \\
& \qquad \left. - \frac{2 \left(1-e^{- \alpha_i \nu \rho_i^*} \right) - \alpha_i \nu \rho_i^* e^{- \alpha_i \nu \rho_i^*}}{\nu^3} \right]\,,  
\end{align}
in which $\alpha_i = \sqrt{2 m (1- \delta_i \kappa_i^*)/C_3 \hbar^2}$, and the right-hand side of this inequality is finite.     

It is time to go back to the inequality~(\ref{phik}), and look at the second independent expression there. After repeating the steps, having been done to obtain Eq.~(\ref{i12}) for $\mathtt{I}$, $\mathtt{II}$ takes the following form,
\begin{align}
\mathtt{II} &\leqslant \frac{C_2 \sqrt{C_3} m}{\hbar^2 \sqrt{1-\delta_i \kappa_i^*}} \int_0^{\rho_i^*} d r \, \frac{J_{L_i}(r)}{r} \left[ e^{ - \sqrt{\frac{2 m (1- \delta_i \kappa_i^*)}{C_3 \hbar^2}} \nu_{i*} r } - e^{ - \sqrt{\frac{2 m (1- \delta_i \kappa_i^*)}{C_3 \hbar^2}} \nu r } \right] \,.
\end{align}
Let us again consider the case $-L_i<0$ only. As $\sinh (\sqrt{L_i} r) /r$ is still a monotonically increasing function, it attains its maximum value at $\rho_i^*$. Evaluating it at this value, and calculating the remaining integral thereafter gives
\begin{align}
\mathtt{II} &\leqslant \frac{C_2 C_3 }{(1-\delta_i \kappa_i^*)} \sqrt{\frac{m}{2 \hbar^2}} \frac{\sinh (\sqrt{L_i} \rho_i^*)}{\sqrt{L_i} \rho_i^*} \left[ \frac{1-e^{ - \sqrt{\frac{2 m (1- \delta_i \kappa_i^*)}{C_3 \hbar^2}} \nu_{i*} \rho_i^*}}{\nu_{i*}} - \frac{1-e^{ - \sqrt{\frac{2 m (1- \delta_i \kappa_i^*)}{C_3 \hbar^2}} \nu \rho_i^*}}{\nu} \right] \,,
\end{align}
whose right-hand side is a finite expression as well as the right-hand side of the inequality above, satisfied by $\mathtt{I}$. Thus, the on-diagonal part of the principal operator is shown to be finite, being  bounded from above by the sum of two finite expressions, 
\begin{align}
\Phi_{ii}(-\nu^2) &\leqslant \mathtt{I} + \mathtt{II} < \infty \,.
\end{align}
We will no longer consider the cases $L \geqslant 0$ here since the calculations apparently repeat themselves. It can, in a similar vein, be shown that the model is also finite for those cases.

\section{Variational approach in the general case}\label{intermezzo}

Since we know that the theory as defined in the previous sections is actually finite, we will generalize the variational approach to the ground state wave function in this section. As we will see,  even to find the solutions to the critical point of the energy functional will be much more complicated, and we will not complete the discussion that this is the true minimum.
Assume that we have $N$ embedded submanifolds, let us make the following choice for the ground state wave function,
\begin{align}
\psi_\alpha(\tilde x|A_i)=\sum_i A_i \Big[{\lambda_i\over V(\Sigma_i)}\Big]^{1/2}\int_{\Sigma_i} d\mu(x)\int_0^\infty {dt\over \hbar}  K_t(x, \tilde x) e^{-\alpha t / \hbar} \,,
\end{align} 
where the factor in front is chosen to make various matrices which will be encountered symmetric. From now on we will denote the measure $d\mu (x)$ on the submanifolds by $d\bar \mu$ to emphasize this factor in front, moreover we  drop the summations over discrete indices, hence  the assumption that the repeated ones are always summed from $1$ to $N$. Since the functions are real, we choose a set of real numbers $A_i$ as weights, indeed we expect the ground state wave function to be positive everywhere.
Since the Hamiltonian contains singular interactions it  is more natural to think of it as a quadratic form and hence apply the variational principle. We will take care of the normalization via a Lagrange multiplier, which corresponds to the energy as is well-known:
\begin{align}
\int_{\cal M} d\tilde \mu(\tilde x) \psi^*(\tilde x) \left[-{\hbar^2\over 2m} \nabla^2_{\tilde g} \psi(\tilde x) \right]-\int_{\Sigma_k} d\bar \mu \psi( x) \int_{\Sigma_k} d\bar \mu' \psi(x')
-E(\int_{\cal M} d\tilde \mu(\tilde x) |\psi(\tilde x)|^2 -1) \,.
\end{align}
Before we move on, let us make a digression and justify that the wave function we construct indeed contains $N$ linearly independent functions, therefore, the variational parameters $A_i$ can be independently varied (apart from the normalization which is also taken care of separately by the Lagrange method).  
Let us assume on the contrary that 
there is a relation among these wave functions,
\begin{align}
\beta_i \int_{\Sigma_i} d\bar\mu \int_0^\infty {dt\over \hbar} K_t(x, \tilde x)e^{-\alpha t / \hbar} =0 \,,
\end{align}
for some set of $\beta_i$ which are real. To gain notational simplicity in our discussion, {\it we set $\hbar=1$ till the end of this section}.
 
Let us multiply this equality by a family of functions, 
\begin{align}
\int_0^\infty {ds} K_s(\tilde x, \tilde y)e^{-\alpha s} {s}^{\rho} \,,
\end{align} 
for $\rho\geq 0$ and integrate over $d\tilde \mu(\tilde x)$. Using the convolution property of the heat kernel and the changes of variables;
$u=t+s$ and $v=t-s$, factoring out the $v$ integral by scaling $v \to uv$, we conclude that 
\begin{align} \label{linind}
\beta_i \int_{\Sigma_i} d\bar \mu \int_0^\infty du K_u(x, \tilde y)e^{-\alpha u} u^{\rho+1} =0 \,,
\end{align}
for all $\rho\geq 0$;  in particular this is true  for $\rho=0,1,2,...$.
By adding to this set of conditions,  the original linear dependence equality as well,  we see 
that the function 
\begin{align} 
 g(t|\tilde x) =\beta_i \int_{\Sigma_i} d\bar\mu  K_t(x, \tilde x)e^{-\alpha t} 
,\end{align}
has a Laplace transform with respect to $e^{-\omega t}=\sum_{k=0}^\infty (-\omega t)^k/k!$ which is identically equal to zero. Note that the  function $g(t|\tilde x)$
vanishes exponentially (for compact submanifold case) as $t\to \infty$,  hence summation and integration can be shown to commute justifying our claim. 
This means the function itself should be zero, $g(t|\tilde x)=0$ (for almost all $t$). Let us assume that we choose $\epsilon$ sufficiently small, and construct 
functions $f^\epsilon_{\Sigma_j}(\tilde x)$ which are mainly concentrated in the $\epsilon$-normal neighborhood of $\Sigma_j$. Therefore a  natural candidate would be the set of functions 
$\int_{\Sigma_j} d\bar \mu'  K_\epsilon(\tilde x, x')$.
Subsequently we calculate the convolution integrals, 
\begin{align}
 \beta_i\int_{\mathcal{M}}\int_{\Sigma_i} d\bar \mu K_\epsilon(x,\tilde x) f^\epsilon_{\Sigma_j}(\tilde x)=\beta_i \int_{\Sigma_i} \int_{\Sigma_j} K_{2\epsilon} (x,x')=0 \,.
\end{align}
If we now take the limit $\epsilon\to 0^+$ we see that the heat kernel approaches to a $\delta$-function on the ambient space, after integration over the variable $\tilde x$, and using the convolution property, we find an integral over  the submanifolds labeled as $\Sigma_i$, this  gives  $\delta_{ij}$.
Thus we conclude that $\beta_j=0$ for all $j$.
This technique will be useful to justify that one of the matrices that we work with is invertible,  as we will see shortly.
 
Let us now write the resulting expressions after the variations with respect to $A_i$ are taken:
\begin{align} 
&\Bigg(\int_{\Sigma_i\times\Sigma_j} d\bar \mu d\bar \mu' \int_0^\infty\, dt\, K_t(x, x') e^{-\alpha_* t} \Bigg)A_{*j}+(-E-\alpha) \Bigg(\int_{\Sigma_i\times\Sigma_j} d\bar \mu d\bar \mu' \int_0^\infty\, dt \,t \, K_t(x, x') e^{-\alpha_* t} \Bigg)A_{*j}\\ \nonumber
& -\Bigg(\int_{\Sigma_i\times\Sigma_k} d\bar \mu d\bar \mu' \int_0^\infty\, dt\, K_t(x, x') e^{-\alpha_* t} \Bigg)\Bigg(\int_{\Sigma_k\times\Sigma_j} d\bar \mu d\bar \mu' \int_0^\infty \, dt \, K_t(y, y') e^{-\alpha_* t} \Bigg)A_{*j}=0 \,,
\end{align}
where we use the symmetry under the interchange of indices $i$ and $j$ and denote the critical solutions with a $*$, notice that there is nothing complex here.
The variation under $\alpha$ now takes the form,
\begin{align}
&(E+\alpha)A_{*i}\Bigg(\int_{\Sigma_i\times\Sigma_j} d\bar \mu d\bar \mu' \int_0^\infty \, dt\, t^2\,  K_t(x, x') e^{-\alpha_* t} \Bigg)A_{*j}\\ \nonumber
& +2A_{*i}\Bigg[\Bigg(\int_{\Sigma_i\times\Sigma_k} d\bar \mu d\bar \mu' \int_0^\infty \, dt \, t\, K_t(x, x') e^{-\alpha_* t} \Bigg)\Bigg(\int_{\Sigma_k\times\Sigma_j} d\bar \mu d\bar \mu' \int_0^\infty \, dt \, t \, K_t(x, x') e^{-\alpha_* t} \Bigg)\\ \nonumber
&\ \ \ \ \ \ \ \ \ \ \ \ \ \qquad \qquad -\Bigg(\int_{\Sigma_i\times\Sigma_j} d\bar \mu d\bar \mu' \int_0^\infty \, dt \,t\, K_t(x, x') e^{-\alpha_* t} \Bigg)\Bigg]A_{*j}=0 \,.
\end{align} 
Let us introduce the following matrices,
\begin{align}
& S_{ij}(\alpha_*)=\Bigg(\int_{\Sigma_i\times\Sigma_j} d\bar \mu d\bar \mu' \int_0^\infty \, dt\, t^2\,  K_t(x, x') e^{-\alpha_* t} \Bigg)=S_{ji}(\alpha_*) \,,\\ \nonumber
& L_{ij}(\alpha_*)=\Bigg(\int_{\Sigma_i\times\Sigma_j} d\bar \mu d\bar \mu' \int_0^\infty \, dt\, t \,  K_t(x, x') e^{-\alpha_* t} \Bigg)=L_{ji}(\alpha_*) \,,\\ \nonumber
& K_{ij}(\alpha_*)=\Bigg(\int_{\Sigma_i\times\Sigma_j} d\bar \mu d\bar \mu' \int_0^\infty \, dt\,  K_t(x, x') e^{-\alpha_* t} \Bigg)=K_{ji} \,,\\ \nonumber
&\tilde \Phi_{ij}(\alpha_*)=\delta_{ij} -K_{ij}(\alpha_*) \,.
\end{align}
We note that if we define $D_{ij}=\sqrt{\lambda_i} \delta_{ij}$, then we have the relation,
\begin{align}
\tilde \Phi_{ij}(\alpha_*) =D_{ik} \Phi_{kl}(\alpha_*) D_{kj} \,,
\end{align} 
where $\Phi$ is as defined in the previous sections.
The variational equations can be recast into the following matrix form after setting $E=-\nu^2$ and $ A=(A_{*i})$, and leaving out $\alpha_*$ dependence of the matrices for simplicity:
\begin{align}
& (\nu^2-\alpha_*) A^TSA+2A^TL\tilde \Phi A=0 \,,\\ \nonumber
& K\tilde \Phi A+ (\nu^2-\alpha_*) LA=0
.\end{align}
Let us note an immediate consequence of these matrix equations, we first transpose the last equation and use the symmetry of $L,K$, following this we then multiply from the left by $\tilde \Phi A$ to get
\begin{align}
A^T\tilde \Phi K \tilde \Phi A+ (\nu^2-\alpha_*) A^T L \tilde \Phi A=0
.\end{align}
Then we use the first equation to eliminate the second term, hence find
\begin{align}
2A^T\tilde \Phi K \tilde \Phi A-(\nu^2-\alpha_*) ^2A^T S A=0
.\end{align}
Here, we  remark that $K, L, S$ are positive definite matrices, moreover $K$ is actually an invertible matrix, as we will elaborate  shortly, hence, if $\alpha_*=\nu^2$, we 
should look for the solutions of 
$\tilde \Phi(\alpha_*)A=0$. If we note now $\tilde \Phi=D\Phi (D A)=0$, we may solve for the vector $DA$. This is exactly the condition we found from the resolvent formula, which defines $\alpha_*$ as well as the relative weights of $(DA)_i$(and they  can all be chosen real), by this choice the $\sqrt{\lambda_i}$ factors in front of the wave functions will disappear. 
 
We will now justify that the matrix $K$ is invertible; note that the if $\beta_i$ is a zero mode of $K$, we may compute $\beta^TK\beta$  as 
\begin{align} 
\beta_i \int_{\Sigma_i\times\Sigma_j} d\bar \mu d\bar \mu' \int_0^\infty \, dt\,  K_t(x, x') e^{-\alpha_* t} \beta_j=\int_{\cal M} d\tilde \mu(\tilde x) \Bigg[ {1\over \sqrt{\pi}} \beta_i\int_{\Sigma_i} d\bar \mu \int_0^\infty \, {du\over \sqrt{u}} K_u(x, \tilde x)e^{-\alpha_* u}\Bigg]^2
\end{align} 
This shows that $K$ is a Gram matrix  of the form $K_{ij}=\int_{\cal M} d\tilde \mu(\tilde x) \phi_i (\tilde x)\phi_j(\tilde x)$.
Therefore a zero mode is possible if the functions $\phi_i(\tilde x)$ are linearly dependent (almost everywhere); 
by multiplying $\phi_i(\tilde x)$ with 
\begin{align} 
\int_0^\infty ds \, K_s(\tilde x, \tilde y) s^{\rho-1/2}e^{-\alpha_* s} \,,
\end{align}
and integrating over $\tilde x$ again, we obtain a set of conditions which will be exactly as before written in Eq.~(\ref{linind}), hence by the same reasoning,
we find that $\beta_i=0$. This implies that $K$ is an invertible matrix.
This allows one to solve for $\tilde \Phi A$ in terms of $K^{-1} L A$ and plug it into the other equation; which gives
\begin{align} 
(\nu^2-\alpha_*) ( A^TSA-2A^TLK^{-1} L A)=0 \,.
\end{align}
We will now see that the matrix in parenthesis is positive definite, which implies that the natural solution will be 
$\alpha_*=\nu^2$.
Let us note the following matrix identity,
\begin{align}
\begin{pmatrix}
I & 0 \\ \sqrt{2}LK^{-1} & I
\end{pmatrix}
\begin{pmatrix}
K & -\sqrt{2}L \\ -\sqrt{2}L & S
\end{pmatrix}
\begin{pmatrix}
I &\sqrt{2} K^{-1}L\\ 0 & I\end{pmatrix}= \begin{pmatrix} K & 0\\ 0 & S-2LK^{-1}L \end{pmatrix} \,.
\end{align}

Hence positivity of the matrix on the left can be checked  by the positivity of the matrix on the right. Therefore,   if we calculate 
the expectation value, for a vector  $(0\  A)$, for the matrix on the right,  we see that it is equal to
\begin{align} 
A^TSA-\sqrt{2} B^TLA-\sqrt{2} A^TLB+B^TKB \,,
\end{align}
where $B=\sqrt{2} K^{-1}LA$, and note that $A^TLB=B^TLA$ since $L^T=L$.
We use now the generalization of the inequality proved in the single submanifold case,
following the same derivation and keeping terms of the form $A_{*i}\int_{\Sigma_i}d\bar \mu$ together, we get
\begin{align} 
\sqrt{2} A^TLB\leq (A^TSA)^{1/2}(B^TKB)^{1/2} \,,
\end{align}
since all the matrices are positive definite, square roots are well defined. We, then, reach to the conclusion that 
\begin{align} 
(\nu^2-\alpha_*) (A^TSA-2A^TLK^{-1}LA)\geq (\nu^2-\alpha_*)[(A^TSA)^{1/2} - \sqrt{2} (A^TLK^{-1}LA)^{1/2}]^2\geq 0 \,.
\end{align}
Hence, a natural solution will  be to take  $\alpha_*=\nu^2$.
\section{Existence of a lower bound for a unique ground state energy}\label{s6}
Following Refs.~\cite{bt1,bt2} we would like to observe the fact that the ground state is bounded from below. It is, however, worth noting the uniqueness of the ground state first. As being done in the aforesaid references, one can obtain a flow equation for each eigenvalue of the principal operator by the well-known Feynman-Hellman theorem~\cite{fh}. Similarly, we here see that the  derivative of the principal operator $\Phi_{ij}(E)$ with respect to $E$ is a strictly negative expression. This suggests that all the eigenvalues of this operator must be monotonically decreasing  functions of $E$. It is, therefore, asserted that {\it the zero of the lowest eigenvalue of the principal operator must correspond to the ground-state energy}.
In these references, by alluding to the Perron-Frobenius theorem, we could prove the nondegeneracy of the ground state. This approach can be repeated verbatim in our case as well. Recall that the ground-state wave function is exactly known due to the discussion in Section~\ref{s1}, and it is immediate to observe that the ground-state is positive. Thus, we reaffirm that the ground-state energy is unique .

The main strand of the method to achieve a lower bound for the ground state energy,  involves the well-known Ger\u{s}gorin theorem~\cite{mat}. According to this theorem, all the eigenvalues $\omega$ of the principal operator as a matrix $\Phi_{ij}(E) \in M_N$ are located in the union of $N$ disks:
\begin{align}
\bigcup_{i=i}^N \vert \omega  - \Phi_{ii} \vert \leqslant \bigcup_{i=i}^N \sum_{i \neq j =1}^N \vert \Phi_{ij} \vert \,.
\end{align}
To utilize this theorem, one should reformulate it in the following manner: If the ground-state energy $E_{gr}$ is bounded from below by some critical value $E_*$, then the eigenvalue $\omega_{min}$ can no longer become zero beyond this bound,  whereby none of the disks can contain $0$ when we set $E$ below $E_*$. It is, therefore, necessary to impose that the principal matrix satisfy the following condition
\begin{align}\label{gg}
\vert \Phi_{ii}(E) \vert^{\min} \geqslant (N-1) \vert \Phi_{ij}(E) \vert^{\max} \,.
\end{align}                  
By estimating the right-hand side from below and the left-hand side from above, and imposing the same inequality for  these estimates, will give us the desired critical bound $E_*$ on the ground state energy.

In the first instance, our main focus is on searching for a lower bound on the on-diagonal part of the principal operator. As  in the search for the lower bound on the coupling constant, the cases here, which involves either an ambient manifold with a non-negative Ricci curvature or  one with a negative Ricci curvature, can be analyzed separately. In this instance, we only consider the case that the ambient manifold is assumed to have a non-negative Ricci curvature, however. We also prefer to concentrate all our efforts on drawing attention to the crucial steps regarding our aim here, rather then repeating the ones having been done before in similar calculations in the previous sections. For simplicity of our presentation we will again assume that the sectional curvature satisfies the bounds $(1-\varepsilon)H\leq Sec\leq H$. As in the previous case there are other possibilities which should be investigated in future work.

Since the ambient manifold is assumed to be a Riemannian manifold with $Ric \geqslant (n-1) K$, the heat kernel bound given in the inequality~(\ref{lkp}) should be applied to the principal operator. As soon as calculating the $t$-integral after plugging this bound into Eq.~(\ref{phi}), we obtain
\begin{align}
\vert \Phi_{ii}(-\nu^2) \vert &\geqslant \frac{m}{2 \pi \hbar^2V(\Sigma_i)} \iint _{\Sigma_i \times \Sigma_i} d \mu(x) d \mu(x') \, \frac{\left[ e^{- \sqrt{\frac{2 m}{\hbar^2}} \nu_{i*} d_\mathcal{M}(x,x')} - e^{- \sqrt{\frac{2 m}{\hbar^2}} \nu d_\mathcal{M}(x,x')}\right]}{d_\mathcal{M}(x,x')} \,.
\end{align}
Here we choose $\nu \in [\nu_{i*},\infty)$. By means of  the same geometric argument as used previously, and applying the Bishop-G\"{u}nter volume comparison theorem, the on-diagonal part of the principal operator takes the following form,
\begin{align}
\vert \Phi_{ii}(-\nu^2) \vert &\geqslant \frac{m}{\hbar^2} \int_0^{\rho_{i*}} d r \, J_{H_i}(r) \frac{\left( e^{- \sqrt{\frac{2 m}{\hbar^2}} \nu_{i*} r} - e^{- \sqrt{\frac{2 m}{\hbar^2}} \nu r } \right)}{r} \,.
\end{align}   
Recall here that $\rho_{i*}=\min_\theta \rho_i(\theta)$ such that $\rho_{i*} \leqslant conv.rad(\Sigma_i)$. Let us analyze three distinct cases due to the specific values of their sectional curvatures:

{\large \textbf{Case $(H_i=0)$}} Our expression is merely an elementary integral, and it gives
\begin{align}
\vert \Phi_{ii}(-\nu^2) \vert & \geqslant \sqrt{\frac{m}{2 \hbar^2}} \left[ \frac{1-e^{-\sqrt{\frac{2 m}{\hbar^2}} \nu_{i*} \rho_{i*}}}{\nu_{i*}} - \frac{1-e^{-\sqrt{\frac{2 m}{\hbar^2}} \nu \rho_{i*}}}{\nu} \right] \,.
\end{align}
The right-hand side of the above inequality is an increasing function since $\nu+1 \leqslant e^\nu$, and it is saturated as $\nu \rightarrow \infty$.
 
{\large \textbf{Cases $(H_i>0)$}} The on-diagonal part of the principal operator is given by 
\begin{align}
\vert \Phi_{ii}(-\nu^2) \vert &\geqslant \frac{m}{\hbar^2} \int_0^{\rho_{i*} \sqrt{H_i}} d \xi \, \frac{\sin \xi}{\sqrt{H_i}}\frac{\left( e^{- \sqrt{\frac{2 m}{\hbar^2 H_i}} \nu_{i*} \xi} - e^{- \sqrt{\frac{2 m}{\hbar^2 H_i}} \nu \xi } \right)}{\xi} \,.
\end{align}
We will apply a slight generalization of the Steffensen's integral inequality~\cite{mit} to this integral so as to estimate it. It states that if $f(\xi)$ is a non-negative monotonically decreasing integrable function on the interval $[a,b]$, and $g(\xi)$ is another integrable function on the same interval such that $0 \leqslant g(\xi) \leqslant A$ for $A>0$, then  
\begin{align}
\int_a^b d \xi \, f(\xi) g(\xi) &\geqslant A \int_{b-\Lambda}^b d \xi \, f(\xi) \,,
\end{align}
where
\begin{align}
\Lambda &= \frac{1}{A} \int_a^b d \xi \, g(\xi) \,.
\end{align}
If this inequality is employed, then one obtains
\begin{align}
\vert \Phi_{ii}(-\nu^2) \vert &\geqslant \frac{m}{\hbar^2} \frac{\sin (\sqrt{H_i} \rho_{i*})}{\sqrt{H_i}} \int_{\rho_{i*} \sqrt{H_i} - \tan \frac{\sqrt{H_i} \rho_{i*}}{2}}^{\rho_{i*} \sqrt{H_i}} d \xi \, \frac{\left( e^{- \sqrt{\frac{2 m}{\hbar^2 H_i}} \nu_{i*} \xi} - e^{- \sqrt{\frac{2 m}{\hbar^2 H_i}} \nu \xi } \right)}{\xi} \,.
\end{align}
Note that $\rho_{i*} \sqrt{H_i} - \tan \frac{\rho_{i*} \sqrt{H_i}}{2} >0$ since $\rho_{i*} \leqslant conv.rad(\Sigma_i)$ is assumed, thereby leading to a non-negative lower limit for the integral. At this point, we should take a very cautious approach while estimating the above integral further. Even if its integrand is an increasing function of $\nu$, it is, at the same time, a decreasing function of the integration variable $\xi$. Thus, any estimation to be made over the integration variable $\xi$ should not conflict with the integrand's monotonicity in the variable $\nu$. Fortunately for us, this integrand, due to being a convex function, allows us to estimate its integral by the Hermite-Hadamard inequality. As a consequence, its monotonicity in the variable $\nu$ remains intact.

Now we make a short digression to justify the convexity of the integrand. It is easy to observe that it is sufficient to consider the following function for the study of the convexity of the integrand,
\begin{align}
\frac{e^{-\xi} - e^{- \lambda \xi}}{\xi} \text{ for } \lambda \geqslant 1\,.
\end{align}
The convexity will amount to the inequality,
\begin{align}
\frac{1}{2} \left[ \frac{e^{-x} - e^{- \lambda x}}{x} + \frac{e^{-y} - e^{- \lambda y}}{y} \right] & \geqslant \frac{e^{-\frac{x+y}{2}} - e^{-\lambda\frac{x+y}{2}}}{\frac{x+y}{2}} \,,
\end{align}
which is equivalent to
\begin{align}\label{inq}
(x+y)\left(y e^{-x} + x e^{-y}\right) - 4 x y e^{- \frac{x+y}{2}} \geqslant (x+y)\left(y e^{-\lambda x} + x e^{- \lambda y}\right) - 4 x y e^{- \lambda \frac{x+y}{2}} \,,
\end{align}
since $x,y > 0$. We consider the following expression for given positive $x$ and $y$ as a function of $\lambda$,
\begin{align}
(x+y) \left( y e^{-\lambda x} + x e^{-\lambda y}\right) - 4 x y e^{- \lambda \frac{x+y}{2}} \,.
\end{align}
The derivative of this function with respect to $\lambda$ is equal to
\begin{align}
2 x y (x+y) \left( e^{- \lambda \frac{x+y}{2}} - \frac{e^{- \lambda x} + e^{- \lambda y}}{2}\right) \,,
\end{align}
which is strictly negative due to the convexity of $e^{- \lambda x}$ in $x$. So the value of this function at $\lambda =1$ is always larger than its value at $\lambda > 1$, which implies the inequality~(\ref{inq}), hence the convexity of the integrand. 

The on-diagonal part of the principal operator takes then the following form,
\begin{align}
\vert \Phi_{ii}(-\nu^2) \vert &\geqslant \frac{m}{\hbar^2} \frac{\left[1- \cos (\sqrt{H_i} \rho_{i*})\right]}{\sqrt{H_i}} \left[ \frac{e^{- \sqrt{\frac{2 m}{\hbar^2 H_i}} \nu_{i*} \left( \rho_{i*} \sqrt{H_i} - \frac{1}{2} \tan \frac{\rho_{i*} \sqrt{H_i}}{2} \right)}}{\rho_{i*} \sqrt{H_i} - \frac{1}{2} \tan \frac{\rho_{i*} \sqrt{H_i}}{2}} \right. \nonumber \\
& \qquad \left. - \frac{e^{- \sqrt{\frac{2 m}{\hbar^2 H_i}} \nu \left( \rho_{i*} \sqrt{H_i} - \frac{1}{2} \tan \frac{\rho_{i*} \sqrt{H_i}}{2} \right)}}{\rho_{i*} \sqrt{H_i} - \frac{1}{2} \tan \frac{\rho_{i*} \sqrt{H_i}}{2}} \right] \,.
\end{align}

{\large \textbf{Case $(-H_i<0)$}} We will only state the result for this case since the whole calculations are the same as the ones in the previous case except that $\sin$ and $\tan$ functions should, respectively, be replaced by $\sinh$ and $\tanh$ functions. Note that the condition $\rho_{i*} \sqrt{H_i} - \tanh \frac{\rho_{i*} \sqrt{H_i}}{2}\geqslant 0$ is always satisfied in this case. Following the previous calculations step by step results in
\begin{align}
\vert \Phi_{ii}(-\nu^2) \vert &\geqslant \frac{m}{\hbar^2} \frac{\left[- 1 + \cosh (\sqrt{H_i} \rho_{i*}) \right]}{\sqrt{H_i}} \left[ \frac{ e^{- \sqrt{\frac{2 m}{\hbar^2 H_i}} \nu_{i*} \left( \rho_{i*} \sqrt{H_i} - \frac{1}{2} \tanh \frac{\rho_{i*} \sqrt{H_i}}{2} \right)}}{\rho_{i*} \sqrt{H_i} - \frac{1}{2} \tanh \frac{\rho_{i*} \sqrt{H_i}}{2}} \right. \nonumber \\
& \qquad - \left. \frac{e^{- \sqrt{\frac{2 m}{\hbar^2 H_i}} \nu \left( \rho_{i*} \sqrt{H_i} - \frac{1}{2} \tanh \frac{\rho_{i*} \sqrt{H_i}}{2} \right)}}{\rho_{i*} \sqrt{H_i} - \frac{1}{2} \tanh \frac{\rho_{i*} \sqrt{H_i}}{2}} \right] \,.
\end{align}
Note that $\vert \Phi_{ii}(-\nu^2) \vert $ is an increasing function of the parameter $\nu$ in all three of the cases as they should be.

It is time to focus on the right-hand side of the inequality~(\ref{gg}). We now need to obtain an upper bound for the off-diagonal part of the principal operator whereby we place the off-diagonal upper bound of the heat kernel~(\ref{ch}) into the off-diagonal part of Eq.(\ref{phi0}). We remark that $\vert \Phi_{ij} \vert \rightarrow 0$ for $i \neq j$ as $\nu \rightarrow \infty$ due to the Lebesgue monotone convergence theorem. After the integration over the time variable $t$, absolute value of this part becomes
\begin{align}\label{pij0}
\vert \Phi_{ij}(-\nu^2) \vert &\leqslant \frac{1}{\sqrt{V(\Sigma_i) V(\Sigma_j)}} \iint_{\Sigma_i \times \Sigma_j} d \mu (x) d \mu (x') \, e^{-\sqrt{\frac{2 m}{C_3 \hbar^2}} \nu d_{\mathcal{M}}(x,x')} \nonumber \\ 
& \qquad \times \left[ \frac{C_1}{V(\mathcal{M})} \left( \sqrt{\frac{2 m}{C_3 \hbar^2}} \frac{d_{\mathcal{M}}(x,x')}{\nu}  + \frac{1}{\nu^2} \right) + \frac{C_2 \sqrt{C_3}}{2 \pi} \frac{m}{\hbar^2 d_\mathcal{M}(x,x')}\right] \,.
\end{align}
The right-hand side is enlarged by introducing a minimum distance between the submanifolds,
\begin{align}
d_{ij} &= \min_{x \in \Sigma_i,x' \in \Sigma_j} d_\mathcal{M} (x,x')\,.
\end{align}
The inequality~(\ref{pij0}), by the definition above, turns into
\begin{align}\label{pij1}
\vert \Phi_{ij}(-\nu^2) \vert &\leqslant \sqrt{V(\Sigma_i) V(\Sigma_j)} \left[ \frac{C_1}{V(\mathcal{M})} \left( \sqrt{\frac{2 m}{C_3 \hbar^2}} \frac{d_{ij}}{\nu} + \frac{1}{\nu^2} \right) + \frac{C_2 \sqrt{C_3}}{2 \pi} \frac{m}{\hbar^2 d_{ij}}\right] e^{-\sqrt{\frac{2 m}{C_3 \hbar^2}} \nu d_{ij}} \,,
\end{align}
which is apparently a decreasing function of the variable $\nu$ as it should be.

Let us study the inequality~(\ref{gg}) for submanifolds with $H_i>0$. The left-hand side of this inequality is supposed to be maximized while the right-hand side thereof being minimized. Introducing the following parameters $\nu^*=\max_i \nu_{i*}$, $\rho_*=\min_i \rho_{i*}$, $\rho^*=\max_i \rho_i^*$, $H_*=\min_i H$, $H^*=\max_i H$, $V^*=\max_i V(\Sigma_i)$, $d_*=\min_{ij} d_{ij}$, and $d^*=\max_{ij} d_{ij}$ obviously allows us to obtain a uniform inequality for the submanifolds that are separated by finite distances. It is as follows  
\begin{align}\label{ggh}
& \frac{m}{\hbar^2} \frac{\left[1- \cos (\sqrt{H_*} \rho_*)\right]}{\sqrt{H^*}} \left[ \frac{e^{- \sqrt{\frac{2 m}{\hbar^2 H_*}} \nu^* \left( \rho^* \sqrt{H^*} - \frac{1}{2} \tan \frac{\rho^* \sqrt{H^*}}{2} \right)}}{\rho^* \sqrt{H^*} - \frac{1}{2} \tan \frac{\rho^* \sqrt{H^*}}{2}} - \frac{e^{- \sqrt{\frac{2 m}{\hbar^2 H_*}} \nu \left( \rho^* \sqrt{H^*} - \frac{1}{2} \tan \frac{\rho^* \sqrt{H^*}}{2} \right)}}{\rho^* \sqrt{H^*} - \frac{1}{2} \tan \frac{\rho^* \sqrt{H^*}}{2}} \right] \nonumber \\
& \qquad \geqslant (N-1) V^* \left[ \frac{C_1}{V(\mathcal{M})} \left( \sqrt{\frac{2 m}{C_3 \hbar^2}} \frac{d^*}{\nu} + \frac{1}{\nu^2} \right) + \frac{C_2 \sqrt{C_3}}{2 \pi} \frac{m}{\hbar^2 d_*}\right] e^{-\sqrt{\frac{2 m}{C_3 \hbar^2}} \nu d_*} \,.
\end{align}
Note that the left-hand side of the inequality~(\ref{ggh}) is an increasing function of $\nu$ whilst the right-hand side is a decreasing function of the same variable, which apparently infers that a solution of this inequality always exists. Thus, it is asserted that the ground state energy is bounded from below, $E_{gr} \geqslant E_*$. The existence of a lower bound on the ground-state energy is self-evident for other cases in which the submanifolds have non-positive sectional curvatures.          

{\large \textbf{Remark.}} At first glance the right-hand side of the inequality~(\ref{ggh}) diverges as the minimum distance $d_*$ goes to 0, which is basically the limit when two submanifolds touch each other. However, this consequence is an immediate aftermath of the way how we estimate the off-diagonal part of the principal operator. It can easily be observed that the off-diagonal part of the principal operator, by Cauchy-Schwarz inequality and the semi-group property of the heat kernel, should satisfy the following condition,  
\begin{align}\label{pij2}
\Phi_{ij}(-\nu^2) &\leqslant \int_0^\infty \frac{d t}{\hbar} e^{- \nu^2 t/\hbar} \left[ \frac{1}{V(\Sigma_i)} \iint_{\Sigma_i \times \Sigma_i} d \mu(x) d \mu(x') \, K_t(x,x') \right]^{1/2} \nonumber \\
& \qquad \times \left[ \frac{1}{V(\Sigma_j)} \iint_{\Sigma_j \times \Sigma_j} d \mu(x) d \mu(x') \, K_t(x,x') \right]^{1/2} \,,
\end{align}
which is  a perfectly well-defined expression, thanks to our discussion on the finiteness of the theory. Therefore, aforementioned divergence is actually an artifact, that is not possessed by the model per se, due to the fact that the model does not even require any renormalization by construction. We would like to demonstrate this fact on a simple, albeit important, example.  

Assume that we have spheres of radii $R_i$, which are isometrically embedded in $\mathbb{R}^3$. Either are they separated by a finite distance, or they are allowed to touch each other. It can easily be shown that after calculating $\Phi_{ii}(-\nu^2)$, the left-hand side of the inequality~(\ref{gg}) attains its minimum value by appropriately substituting the parameters $R_{\min} =\min_i R_i$, $R_{\max} =\max_i R_i$, $\nu_{\min} =\min_i \nu_{i*}$, and $\nu_{\max} =\max_i \nu_{i*}$ for $R_i$ and $\nu_{i*}$. 
\begin{align}
\vert \Phi_{ii}(-\nu^2) \vert^{\min} &\geqslant \sqrt{\frac{2 m}{\hbar^2}} \left[ \frac{1-e^{- \sqrt{\frac{2 m R_{\min}^2}{\hbar^2}} 2 \nu_{\min}}}{2 \nu_{\max}} - \frac{1-e^{- \sqrt{\frac{2 m R_{\max}^2}{\hbar^2}} 2 \nu}}{2 \nu} \right] \,.
\end{align}

The off-diagonal part of the principal operator can be estimated by the inequality~(\ref{pij2}). Its absolute value attains a maximum value via replacing the parameters $R_i$ and $\nu_{i*}$ by the ones introduced in the previous paragraph, and it is given by
\begin{align}
\vert \Phi_{ij} (-\nu^2) \vert^{\max} &\leqslant \sqrt{\frac{2 m}{\hbar^2}} \left[ \frac{1-e^{- \sqrt{\frac{2 m R_{\max}^2}{\hbar^2}} 2 \nu}}{ 2 \nu} + \frac{e^{- \sqrt{\frac{4 m R_{\min}^2}{\hbar^2}} 2 \nu}}{8 \nu} \right] \,.
\end{align}  
Thus, the inequality~(\ref{gg}) takes the following form,
\begin{align}
\frac{1-e^{- \sqrt{\frac{2 m R_{\min}^2}{\hbar^2}} 2 \nu_{\min}}}{\nu_{\max}} - \frac{1-e^{- \sqrt{\frac{2 m R_{\max}^2}{\hbar^2}} 2 \nu}}{\nu} &\geqslant (N-1) \left[ \frac{1-e^{- \sqrt{\frac{2 m R_{\max}^2}{\hbar^2}} 2 \nu}}{\nu} + \frac{e^{- \sqrt{\frac{4 m R_{\min}^2}{\hbar^2}} 2 \nu}}{4 \nu} \right] \,,
\end{align}
which is completely insensitive to whether spheres are apart or touch each others. 
\section{Comments on more general problems}\label{s7}
It is interesting that the resolvent approach which involves the principal matrix $\Phi$ becomes indispensable when we have singular interaction supported on surfaces as well as supported on curves  and points. For simplicity we will discuss a single  surface $\Sigma$ and a single point source, located at some point $a$,  more general cases can be worked out in a similar way.
We will assume that the surface interaction is specified by a coupling constant which may or may not lead to a bound state, yet the point has  a bound state of energy $-\mu^2$, and its interaction in general can not be defined by a finite coupling constant. 
In this case, going over the resolvent, and introducing the principal matrix as before we arrive the same resolvent formula, as spelled out in Section~\ref{s1}, and the matrix $\Phi$ becomes,
\begin{align}
\begin{pmatrix}
{1\over \lambda}-{1\over V(\Sigma)} \int_{\Sigma\times \Sigma}d\mu(x)d\mu(x')\int_0^\infty {dt\over \hbar}K_t(x,x')e^{-{\nu^2t / \hbar}} & -{1\over \sqrt{V(\Sigma)}}\int_{\Sigma}d\mu(x)\int_0^\infty {dt\over \hbar} K_t(x, a)e^{-{\nu^2t / \hbar}}
 \\    -{1\over \sqrt{V(\Sigma)}}\int_{\Sigma}d\mu(x)\int_0^\infty {dt\over \hbar} K_t(x, a)e^{-{\nu^2t / \hbar}} & \int_0^\infty {dt\over \hbar} K_t(a,a)(e^{-{\mu^2 t / \hbar}}-e^{-{\nu^2t / \hbar}}) \end{pmatrix} \,.
\end{align}
The zero eigenvalues of this matrix  determines the possible bound state energies of this combined system.
By the Cauchy interlacing theorem of matrices, we may assert that the lowest eigenvalue of this matrix will always cut the real axis at a value lower than $-\mu^2$, hence assuring the existence of a bound state. The analysis of the possible appearance of a second bound state is not simple and depends on the details of the geometry of this configuration.

An interesting problem is to consider a point which is located sufficiently far away from the surface in a noncompact ambient manifold ${\mathcal M}$. This is equivalent to the following condition: we denote the minimum distance between the surface and the point by $d_*$, which exists since the surface $\Sigma$ is compact, then we suppose that  ${\hbar^2\over 2m d_*^2}<<\mu^2$. The same condition applies if the surface is assumed carry a bound state solution. In this case one can see that the off-diagonal elements satisfy, within the given approximation,
\begin{align}
\Big\vert{1\over \sqrt{V(\Sigma)}}\int_{\Sigma}d\mu(x)\int_0^\infty {dt\over \hbar} K_t(x, a)e^{-{\mu^2t\over \hbar}}\Big\vert\leqslant 
C_1{\sqrt{V(\Sigma)}\over d_*} e^{-C_2{\sqrt{2m}d_*\mu\over\hbar}} \,,
\end{align}
where $C_1$ and  $C_2<1/2$ are some constants. This shows that the off-diagonal elements become small perturbations.
Hence a perturbation analysis, which is very similar to the one worked out in \cite{et}, shows that the shift in the bound state energy $\mu^2$  is given by 
\begin{align}
& \delta \mu^2=
\Big|{1\over \sqrt{V(\Sigma)}}\int_{\Sigma}d\mu(x)\int_0^\infty {dt\over \hbar} K_t(x, a)e^{-{\mu^2t\over \hbar}}\Big|^2\nonumber\\
&\ \ \ \ \ \ \ \ \times \Big[\int_0^\infty {dt\over \hbar} {t\over \hbar} K_t(a,a)e^{-{\mu^2 t\over \hbar}}\Big]^{-1} \Big[{1\over \lambda}-{1\over V(\Sigma)} \int_{\Sigma\times \Sigma}d\mu(x)d\mu(x')\int_0^\infty {dt\over \hbar}K_t(x,x')e^{-{\mu^2t\over  \hbar}} \Big]^{-1} 
,\end{align}
which is exponentially small as it should be for a tunneling solution. Note that this solution breaks down if the surface has a bound state exactly at the same value $-\mu^2$, which is not surprising since then we must resort to degenerate perturbation theory to get the solution. 
Similar comments will apply for a family of curves and surfaces supporting singular interactions. We plan to come back to some of these issues in future work.

\section{Acknowledgment}
This work is supported by Bo\u{g}azi\c{c}i University BAP Project \#:6513.

\end{document}